\documentclass{mhd}
\pdfoutput=1
\usepackage{natbib}
\bibpunct{(}{)}{;}{a}{}{,} 
\usepackage[pdftex]{graphicx}
\usepackage{txfonts}
\usepackage{afterpage}

\usepackage{lscape} 

\begin{document}
\title{Characterizing star formation activity in infrared dark cloud MSXDC G048.65-00.29}
\titlerunning{Characterizing Star Formation Activity in IRDC G48.65}

\author{M.~H.~D.~van der Wiel\inst{1,2} \and R.~F.~Shipman\inst{2,1}}
\authorrunning{M.~H.~D.~van der Wiel \and R.~F.~Shipman}
\institute{
Kapteyn Astronomical Institute, PO Box 800, 9700 AV, Groningen, The Netherlands\\email: {\tt wiel@astro.rug.nl}
\and 
SRON Netherlands Institute for Space Research, PO Box 800, 9700 AV, Groningen, The Netherlands
}

\date{Received 3 March 2008 / Accepted 13 July 2008}

\abstract
{Infrared Dark Clouds (IRDCs), condensed regions of the ISM with high column densities, low temperatures and high masses, are suspected sites of star formation. Thousands of IRDCs have already been identified. To date, it has not been resolved whether IRDCs always show star formation activity and, if so, if massive star formation ($\gtrsim8 M_\odot$) is the rule or the exception in IRDCs.}
{Previous analysis of sub-millimeter cores in the cloud MSXDC G048.65$-$00.29 (G48.65) indicates embedded star formation activity. To characterize this activity in detail, mid-infrared photometry (3--30~$\mu$m) has been obtained with the \textit{Spitzer Space Telescope}. This paper analyzes the point sources seen in the 24~$\mu$m band, combined with counterparts or upper limits at shorter and longer wavelengths.}
{Data points in wavelength bands ranging from 1~$\mu$m up to $850~\mu$m (\textit{Spitzer} IRAC, MIPS~24 and 70~$\mu$m, archival 2MASS data and sub-millimeter counterparts) are used to compare each 24~$\mu$m source to a set of Spectral Energy Distributions of Young Stellar Object (YSO) models. By assessing the models that fit the data, an attempt is made to identify YSOs as such and determine their evolutionary stages and stellar masses.}
{A total of 17 sources are investigated, 13 of which are classified as YSOs, primarily -- but not exclusively -- in an early embedded phase of star formation. The modeled masses of the central stellar objects range from sub-solar to $\sim$8 $M_\odot$. Every YSO is at less than 1 pc projected distance from its nearest YSO neighbor.}
{IRDC G48.65 is a region of active star formation. We find YSOs in various evolutionary phases, indicating that the star formation in this cloud is not an instantaneous process. The inferred masses of the central objects suggest that this IRDC hosts only low to intermediate mass YSOs and none with masses exceeding $\sim$8 $M_\odot$.}


\maketitle


\section{Introduction}

Infrared dark clouds (IRDCs), discovered independently by \citet{perault1996} and \citet{egan1998} as dark patches against the Galactic mid-infrared background, are suspected to be sites of clustered star formation \citep{rathborne2006,simon2006}. Moreover, the suggestion is raised \citep[see e.g.][]{menten2005,beuther2007} that IRDCs harbor massive young stars. Morphologies of IRDCs range from globular to more filamentary structures. One of the most important properties of IRDCs is their high column density \citep[$\gtrsim10^{22}\ \mathrm{cm^{-2}}$, e.g.][]{carey2000}. Dust absorbs mid-infrared radiation and makes the clouds stand out as dark features against the mid-infrared background. Temperatures in IRDCs are $< 25 \ \mathrm{K}$ and volume densities can exceed $10^5 \ \mathrm{cm^{-3}}$ \citep{egan1998,carey2000}. Typical size and mass scales are $\sim$5 pc and $10^3$--$10^4~M_\odot$ \citep{simon2006}. These low initial temperatures, high molecular densities and total masses typical for IRDCs are precisely what make them suitable candidates for star forming regions. IRDCs are generally not quiescent: they harbor compact cores of sub-millimeter emission \citep[e.g.][]{carey2000,ormel2005,rathborne2005}. IRDCs are found primarily in the inner Galaxy and close to the Galactic plane. Thousands are known in the first and fourth quadrants of the Galactic plane \citep{simon_catalog2006}, i.e., the inner Galaxy. \citet{frieswijk2007} recently identified the first IRDC in the outer Galaxy (in the absence of a bright mid-infrared background), using 2MASS (Two Micron All Sky Survey) color distributions of background stars, followed up by observations of molecular lines and \textit{Spitzer Space Telescope} photometry. 

It has not yet been established whether all IRDCs show active star formation or that some may in fact harbor only starless cores. Moreover, the association of IRDCs to \emph{massive} ($\gtrsim8 M_\odot$) star formation in particular is still a matter of dispute. Studies of star forming regions in general \citep[e.g.][]{indebetouw2007} and IRDCs in particular \citep[e.g.][]{beuther2007} prove that star formation activity in a wide stellar mass range and at various phases of evolution can be probed by Spitzer photometry. 

The IRDC under investigation in this paper is the cloud at Galactic coordinates $(\ell,b)=(48\fdg66, -0\fdg30)$. It has been previously observed by the Mid-course Space Experiment (MSX), the SCUBA instrument on the James Clerk Maxwell Telescope \citep[JCMT,][]{ormel2005}, and by the JCMT in CO, $^{13}$CO and HCO$^+$ \citep{shipman2003}. In addition, it is covered by the Galactic Legacy Infrared Mid-Plane Survey Extraordinaire \citep[GLIMPSE,][]{benjamin2003}. In the dark cloud catalog of \citet{simon_catalog2006} this cloud has the designation ``MSXDC G048.65$-$00.29", hereafter simply ``G48.65".  Its distance is determined kinematically from molecular line data and an assumed Galactic rotation curve. G48.65 is found to be at a distance of $\sim$2.5$\ \mathrm{kpc}$ \citep{ormel2005,simon2006}. Its distance to the Galactic Center is $\sim$7~$\mathrm{kpc}$ and it is less than 20 pc away from the mid-plane of the Galaxy. Its total mass is estimated at almost 600\,$M_\odot$ within a 2 pc area and the molecular (H$_2$) density is $\sim$$10^3 \ \mathrm{cm^{-3}}$ \citep{simon2006}. \citet{ormel2005} identified three distinct emission cores at 450 and 850 $\mu$m; their modeling indicates the presence of central luminosity sources on the order of $10^2$--$10^3 \,L_\odot$ in at least two of these cores. 

The earliest phases of star formation show emission at $>$30$~\mu$m as they are embedded in an accreting envelope; the peak wavelength shifts through the mid- and near-infrared to the optical regime as stars evolve toward the main sequence. As mentioned above, sub-millimeter observations indicate ongoing embedded star formation in two or three cores in G48.65 \citep{ormel2005}. This raises the question whether the star forming activity in G48.65 is limited to the sub-millimeter cores. Additionally, it is realized that observations at shorter wavelengths can improve our understanding of the previously identified star forming cores. This was the motivation to take a closer look in the mid-infrared (roughly 3--30\,$\mu$m), where Young Stellar Objects (YSOs) are known to emit the bulk of their energy. This paper describes the interpretation of \textit{Spitzer Space Telescope} photometry of G48.65, focussing on those sources that show emission at 24\,$\mu$m, where the brightest, massive, embedded young stars are expected to be found. 

Section \ref{sec:observations} describes the Spitzer data and their reduction, Section \ref{sec:results} deals with the classification of twenty sources based on the available data points in the near- and mid-infrared and sub-millimeter regimes. Conclusions and a discussion are presented in Sect.~\ref{sec:conclusions_and_discussion}.

\section{Observations}
\label{sec:observations}

\subsection{Spitzer photometry}
\label{sec:spitzerdata}

Mid-infrared images of IRDC G48.65 were recorded in October 2004 by the Infrared Array Camera (IRAC) and Multiband Imaging Photometer (MIPS) on board the {\it Spitzer Space Telescope}, as part of the proposal by Kraemer et al.\ (General Observing Proposal \#3121). Photometric images were obtained in five broadband filters: all four bands of IRAC (centered around 3.6, 4.5, 5.8 and 8.0\,$\mu$m) and the first band of MIPS (24\,$\mu$m). 

For the IRAC bands, the 12 second High Dynamic Range mode is used: five dithered images of \mbox{10.4 s} and five of \mbox{0.4 s}, adding up to a total integration time of 52+2 seconds per IRAC band per sky position. IRAC observes two adjacent $5\arcmin\times5\arcmin$ fields simultaneously, MIPS observes only one at a time. MIPS images are also dithered, a total of 44 images spread over the two fields add up to an integration time of 55 s per field. The resolution in band 1 of IRAC is limited by the pixel size of the CCD ($\sim$$1\farcs2$); it is diffraction limited for the higher wavelengths, with the resolution element ranging from 1\farcs3 for \mbox{IRAC 4.5\,$\mu$m} to $6\arcsec$ for the \mbox{24\,$\mu$m} MIPS band \citep{rieke2004,fazio2004}.

The Spitzer Science Center (SSC) provides `basic calibrated data' in units of physical flux per unit solid angle. The basic calibrated data originate from pipeline version S11.4.0 in case of IRAC and version S10.5.0 for the MIPS band. These images are processed and combined into mosaics using scripts from the MOPEX software package \citep{makovoz_marleau2005,makovoz_khan2005}. Data from the 0.4 s exposures is used to replace saturated pixels in the 10.4 s images.

\begin{figure} 
	\resizebox{\hsize}{!}{\includegraphics{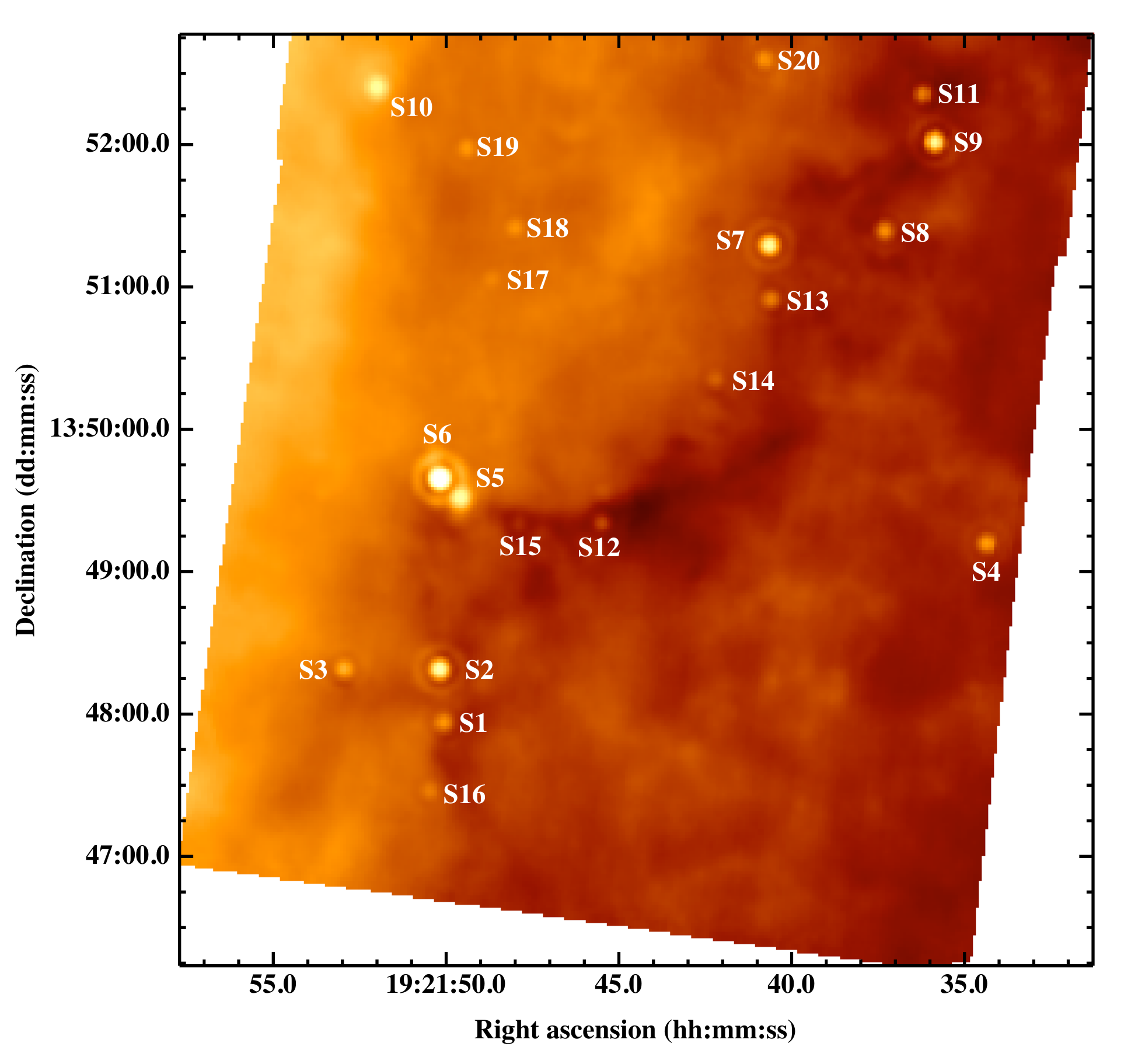}} 
	\caption{Twenty point sources visible in the 24 $\mu$m MIPS image. Axes are in J2000 coordinates. The brightness scale is logarithmic, ranging from 27 (black), through red and yellow, to 170 MJy/sr (white). The ``S" numbers correspond to the source IDs as listed in Table \ref{table:sourcelist}. The annuli around the brightest sources are artefacts of the instruments point spread function.} 
	\label{fig:MIPSsources} 
\end{figure}

\subsection{Point source extraction}
\label{sec:PSE}
At Galactic coordinates $(\ell,b)=(48\fdg66,-0\fdg30)$, IRDC G48.65 lies close to the line of sight toward the giant molecular cloud (GMC) W51. Along a line of sight this close to the Galactic plane in general and the bright GMC in particular, there is significant background radiation in the mid-infrared. The background varies with amplitudes comparable to some of the point source peaks and on size scales smaller than the point spread functions. This complicates (automated) point source extraction, especially at the longer wavelength bands at 8.0 $\mu$m and 24 $\mu$m. In the shorter wavelength bands, spatial variability of the background is less of an issue, but source confusion starts to play a role. 

The MIPS image (at  24\,$\mu$m) shows 20 distinct point source objects (see Fig.~\ref{fig:MIPSsources}). A subpixel position is determined for each of the 20 sources by fitting radial profiles in the 24\,$\mu$m image. Point spread function (PSF) photometry is performed at these positions, using the `optimal photometry' routine \citep{naylor1998} built into the STARLINK \emph{GAIA} tool. The PSF itself is modeled on source S2, being bright and relatively isolated. The PSF is truncated at a radius of 6\farcs1, the sky anulus is defined to be between 12\farcs3 and 15\farcs3. An aperture correction factor of 1.7, appropriate for this radius\footnote{{\tt http://ssc.spitzer.caltech.edu/mips/apercorr/}}, is applied. Sources S12 and S15 require a smaller truncation radius (3\farcs9) in order to prevent overestimating the background sky level; in this case, an aperture correction factor of 2.5 is applied. The uncertainties listed for every 24\,$\mu$m flux value in Table \ref{table:sourcelist} include sky pixel variance and the 4\% uncertainty in the absolute calibration \citep{engelbracht2007}.

\begin{landscape}
\begin{table} 

\begin{minipage}[t]{\textwidth}

\caption{Point source fluxes}
\label{table:sourcelist}

\centering
\renewcommand{\footnoterule}{}  
\tiny  
\begin{tabular}{r c c | r | r r r r r r r r r r r r | r}
\hline\hline
ID & & & 2MASS designation & & & & & \multicolumn{4}{c}{\emph{Spitzer} fluxes and uncertainties\footnote{A ``$<$" sign indicates that the corresponding flux value is an upper limit. The IRAC 3.6~$\mu$m and 4.5~$\mu$m fluxes and uncertainties for S4, S17 and S18 are taken from GLIMPSE.}} & & & & & SCUBA peak\footnote{Sub-millimeter peaks P1, P2 and EP are listed in Table 2 of \citet{ormel2005}. See text in Sec.~\ref{sec:stageIobjects} for details of P1's subdivision. P3 is not defined in \citet{ormel2005}, its flux at 450 $\mu$m is estimated in Sec.~\ref{sec:stageIobjects} of the present study.}\\
 & RA	& Dec  & $J$, $H$ \& $K_s$ bands & 3.6 $\mu$m	& $\sigma_{3.6}$	& 4.5 $\mu$m & $\sigma_{4.5}$ & 5.8 $\mu$m & $\sigma_{5.8}$ & 8.0 $\mu$m & $\sigma_{8.0}$ & 24 $\mu$m & $\sigma_{24}$ & 70 $\mu$m\footnote{The MIPSGAL mosaic shows one emission core of 6.9 Jy at the position of S5 and S6. See Sect.~\ref{sec:stageIobjects} for details of the subdivision of the MIPS 70~$\mu$m flux and Sect.~\ref{sec:SCUBAand2MASS} for the upper limit used for other sources.} & $\sigma_{70}$ & sub-millimeter \\
 &  (J2000) & (J2000) & & (mJy) & (mJy) & (mJy) & (mJy) & (mJy) & (mJy) & (mJy) & (mJy) & (mJy) & (mJy) & (Jy) & (Jy)  \\
\hline
    S1 & 19$^\mathrm{h}$21$^\mathrm{m}$50$\fs$1 & +13$^\circ$47$\arcmin$57$\arcsec$ & - &   $<$ 0.500  &   0.250   &   $<$ 0.500   &  0.250    &  $<$ 2.000  &   1.000 &     $<$ 2.000    & 1.000    & 26.647    &  3.033 & - & - & - \\
    S2  & 19$^\mathrm{h}$21$^\mathrm{m}$50$\fs$2 & +13$^\circ$48$\arcmin$19$\arcsec$ & 19215022+1348185 &   2.994   &   0.518   &   6.248   &   0.854    &  8.012   &   1.115   &   3.862     & 1.107   & 101.585     & 7.051& - & -	& P3   \\
   S3  & 19$^\mathrm{h}$21$^\mathrm{m}$53$\fs$0 & +13$^\circ$48$\arcmin$19$\arcsec$  & - &  $<$ 0.500    & 0.250   &   $<$ 0.500    & 0.250    &  $<$ 2.000    & 1.000    &  $<$ 2.000    & 1.000     & 23.254     & 2.999 & - & - & - \\
    S4  & 19$^\mathrm{h}$21$^\mathrm{m}$34$\fs$4 & +13$^\circ$49$\arcmin$12$\arcsec$ & 19213444+1349122 & 234.100   &   8.137   & 161.000  &  6.024  &  148.499  &   10.147  &   75.078  &   5.725  &   32.944   &   2.963 & - & - & - \\
    S5  & 19$^\mathrm{h}$21$^\mathrm{m}$49$\fs$6 & +13$^\circ$49$\arcmin$31$\arcsec$ & -& 3.741    &  0.622   &  4.808     & 0.720   & 29.020     & 2.581   & 58.549    & 4.542   & 109.226    &  7.936 & 2.4 & 1.4 & part of P1 \\
    S6  & 19$^\mathrm{h}$21$^\mathrm{m}$50$\fs$2 & +13$^\circ$49$\arcmin$39$\arcsec$ & - &$<$ 0.500    & 0.250    &  4.237    & 0.682    &  8.802     & 1.291     & 4.214     & 1.213    & 295.503   &  16.843 & 4.5 & 1.4 & part of P1  \\
    S7  & 19$^\mathrm{h}$21$^\mathrm{m}$40$\fs$7 & +13$^\circ$51$\arcmin$17$\arcsec$ & 19214073+1351176 & 1.233     & 0.314     & 1.581     & 0.352   &   $<$ 2.000    & 1.000     & $<$ 2.000    & 1.000    & 90.409     & 6.485 & - & - & -  \\
    S8  & 19$^\mathrm{h}$21$^\mathrm{m}$37$\fs$3 & +13$^\circ$51$\arcmin$23$\arcsec$ & 19213739+1351238 & 4.448    & 0.662    & 9.188     & 1.140   &  9.272    &  1.200    & 13.316    &  1.768   &  31.222    &  2.877 & - & - & -  \\
    S9  & 19$^\mathrm{h}$21$^\mathrm{m}$35$\fs$9 & +13$^\circ$52$\arcmin$01$\arcsec$ & 19213593+1352007 & 4.001    &  0.653    &  3.637     & 0.609    &  4.461     & 0.779    & 9.529     & 1.468    & 87.157    &  6.167 & - & - & -   \\
    S10  & 19$^\mathrm{h}$21$^\mathrm{m}$52$\fs$1 & +13$^\circ$52$\arcmin$24$\arcsec$  & - &  $<$ 0.500    &  0.250    &  0.639    &  0.196     & $<$ 2.000    & 1.000     & $<$ 2.000    & 1.000    & 67.752    &  5.307 & - & - & - \\
    S11 & 19$^\mathrm{h}$21$^\mathrm{m}$36$\fs$2 & +13$^\circ$52$\arcmin$21$\arcsec$  & - & 0.201    & 0.122     & 1.109    &  0.293     & 1.206   &   0.429     & 1.988     & 0.959    & 24.073     & 2.446 & - & - & -  \\
    S12  & 19$^\mathrm{h}$21$^\mathrm{m}$45$\fs$6 & +13$^\circ$49$\arcmin$21$\arcsec$ & - &  $<$ 0.500    & 0.250     & 0.260    & 0.133     & $<$ 2.000   &  1.000     & $<$ 2.000    & 1.000     & 13.778     & 2.018 & - & - & EP  \\
    S13 & 19$^\mathrm{h}$21$^\mathrm{m}$40$\fs$7 & +13$^\circ$50$\arcmin$55$\arcsec$ &  19214066+1350547 &  4.342    &  0.695     & 4.106    &  0.659     & 2.650     & 0.578     & $<$ 2.000    & 1.000    & 16.289     & 2.513 & - & - & -  \\
    S14 & 19$^\mathrm{h}$21$^\mathrm{m}$42$\fs$3 & +13$^\circ$50$\arcmin$21$\arcsec$ & 19214228+1350212 &  3.269     & 0.589     & 2.934    &  0.526     & 3.189     & 0.594     & 3.125     & 0.746     & 7.466     & 1.673 & - & - & -  \\
    S15  & 19$^\mathrm{h}$21$^\mathrm{m}$47$\fs$9 & +13$^\circ$49$\arcmin$20$\arcsec$ & - &  $<$ 0.500    &  0.250 &   0.726 &  0.222 &  $<$ 2.000 &  1.000  & $<$ 2.000 &  1.000 & 2.757  & 0.918 & - & - & P2 \\
    S16 & 19$^\mathrm{h}$21$^\mathrm{m}$50$\fs$5 & +13$^\circ$47$\arcmin$28$\arcsec$ & - &  3.303    &  1.331     & 2.624     & 1.304     & 6.388     & 0.908     & 2.290     & 0.687     & 5.544    &  1.381 & - & - & -  \\
    S17  & 19$^\mathrm{h}$21$^\mathrm{m}$48$\fs$7 & +13$^\circ$51$\arcmin$02$\arcsec$   & 19214875+1351034 &  212.300  &  7.346  &  111.500  & 4.275   &  81.086   &  6.043   &  41.064    &  3.503   &   7.142  &   1.262 & - & - & -  \\
    S18 & 19$^\mathrm{h}$21$^\mathrm{m}$48$\fs$1 & +13$^\circ$51$\arcmin$24$\arcsec$ & 19214807+1351249 & 436.200    & 18.300   & 240.600    & 14.500   & 87.205  &   6.401   & 87.228  &   6.439  &    13.605  &    1.598 & - & - & - \\
    S19 & 19$^\mathrm{h}$21$^\mathrm{m}$49$\fs$4 & +13$^\circ$51$\arcmin$58$\arcsec$ & 19214945+1351581 & 416.400    & 18.690   & 233.300     & 9.739   & 69.551   &  5.225   & 88.545    & 6.542     & 12.832     & 1.589 & - & - & - \\
    S20 & 19$^\mathrm{h}$21$^\mathrm{m}$40$\fs$8 & +13$^\circ$52$\arcmin$36$\arcsec$ & - & 14.478   & 1.442    & 14.110     & 1.531    & 12.216    & 1.410    & 6.380     & 1.035    & 22.614    &  2.258 & - & - & - \\
\hline
\end{tabular}
\normalsize
\end{minipage}
\end{table}
\end{landscape}

The positions of the 24~$\mu$m sources are used as starting positions for the photometry in the four IRAC bands. We apply the same method as described above, the only differences being:
\begin{itemize}
	\item[(i)] the PSF is modeled on a bright isolated star in each image (S2 is not a suitable candidate at IRAC wavelengths); 
	\item[(ii)] the PSF is truncated at 4\farcs8 for IRAC~3.6 and 4.5 $\mu$m and at 5\arcsec\ for IRAC 5.8 and 8.0~$\mu$m; 
	\item[(iii)] the relevant aperture correction is roughly 1.1; 
	\item[(iv)] the absolute calibration uncertainty is taken to be 5\% \citep{reach2005}.
\end{itemize}

The fluxes of the 20 MIPS sources, their counterparts in the IRAC bands, and the corresponding uncertainties are listed in Table \ref{table:sourcelist}. Sources S4, S17, S18 and S19 suffer from bad pixel columns at the two shortest wavelengths; for these sources, the fluxes and uncertainties at 3.6 and 4.5\,$\mu$m are taken from the GLIMPSE archive. At the positions where no counterpart was found for a MIPS source in a particular IRAC band, an upper limit to the flux value, indicated by a ``$<$" symbol, is listed in Table \ref{table:sourcelist}. Since we want to avoid underestimating upper limits to the flux values in crowded, embedded and extincted regions, we conservatively set the upper limits around once or twice the lowest flux value that \emph{is} detected in the same band. Here, the uncertainties are estimated at 50\% of the flux limit value.

\subsection{Additional data: sub-millimeter, far- and near-infrared}
\label{sec:SCUBAand2MASS}
In addition to the newly obtained \textit{Spitzer Space Telescope} mid-infrared photometry, we use sub-millimeter data points from \citet{ormel2005} obtained with SCUBA at JCMT, MIPSGAL 70\,$\mu$m far-infrared images \citep{carey2006}, and archival near-infrared data from the 2MASS catalog \citep{skrutskie2006}. See the corresponding columns in Table \ref{table:sourcelist} for the objects that have counterparts in the near-infrared (2MASS) or in the sub-millimeter (SCUBA).

\begin{figure}[h]
	\resizebox{\hsize}{!}{\includegraphics{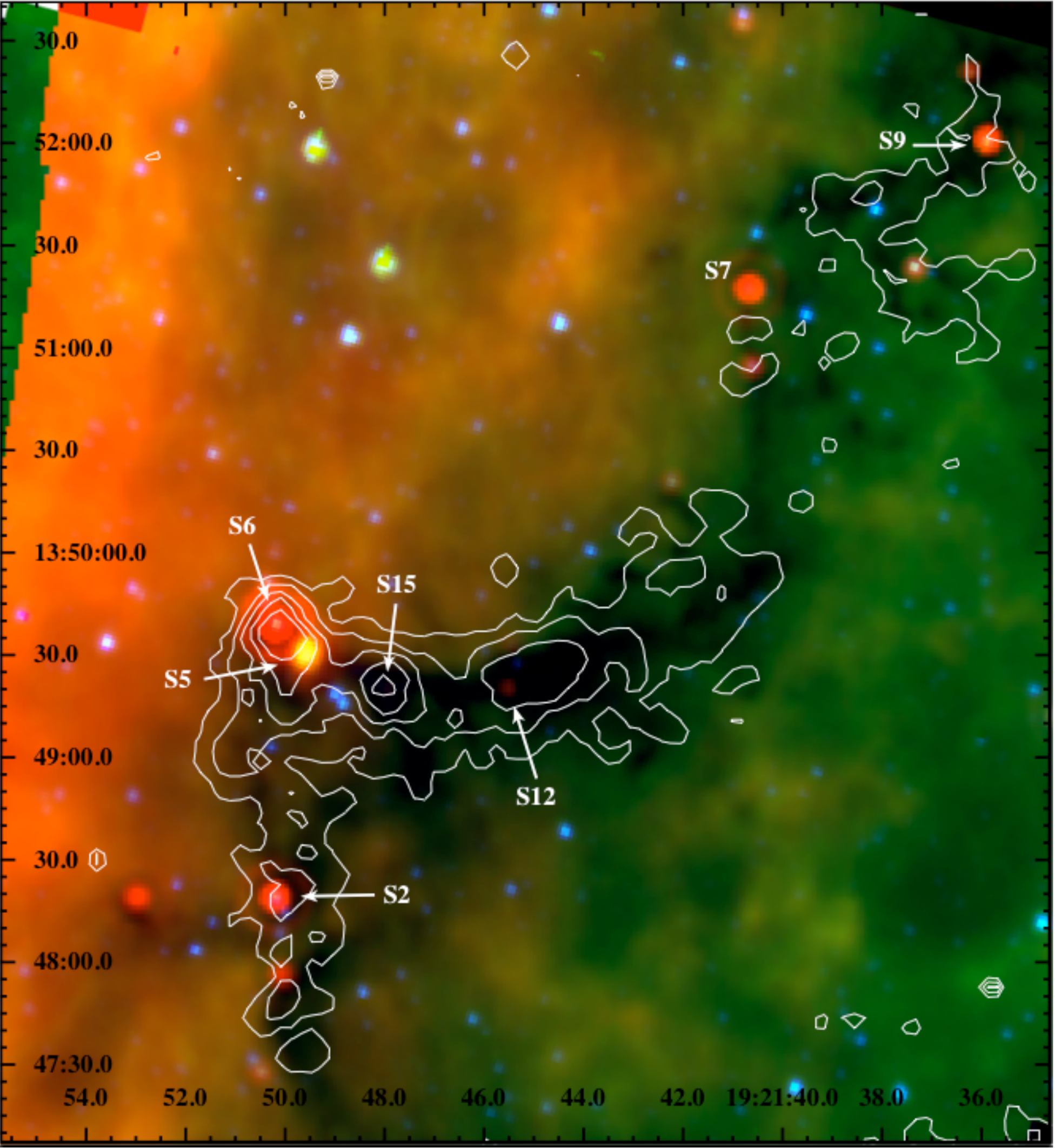}} 
	\caption{Composite image of IRDC G48.65. The color image is composed of images of the Spitzer passbands at 24\,$\mu$m (red), 8.0\,$\mu$m (green) and 4.5\,$\mu$m (blue). Overlaid are the 850\,$\mu$m SCUBA contours at levels of 0.08, 0.16, 0.24, 0.32 and 0.40\,Jy/beam. The axes annotate right ascension and declination (J2000). Notable MIPS point sources are labeled.}
	\label{fig:G48_3color} 
\end{figure}

The SCUBA data point named P1 is matched to S5 and S6 and is discussed in Sect.~\ref{sec:stageIobjects}. The positions of the peaks P2 and EP as listed in Table~2 of \citet{ormel2005} are incorrect and have an offset of $\sim$15$\arcsec$ in RA with respect to the true peak positions in the 850~$\mu$m and 450~$\mu$m images (C.W.~Ormel, priv.~comm., see also Fig.~\ref{fig:G48_3color}). The derived fluxes of the two peaks are nevertheless correct. The correct peak position for P2 is (RA $=$ 19$^\mathrm{h}$21$^{\mathrm{m}}$48\fs0, Dec $=$ $+13^\circ49\arcmin21\arcsec$) ($\pm 2\arcsec$, J2000); the correct peak position for EP is (19$^\mathrm{h}$21$^{\mathrm{m}}$44\fs9, $+13^\circ49\arcmin24\arcsec$) ($\pm 5\arcsec$, J2000). The mid-infrared source listed as S12 in Table~\ref{table:sourcelist} can therefore be matched to EP in the sub-millimeter, and S15 to P2. 

The third brightest 24\,$\mu$m object, S2, is also seen in the 450\,$\mu$m SCUBA image, albeit less bright than the other three peaks. Since it is not in the list of SCUBA peaks of \citet{ormel2005}, we estimate its 450\,$\mu$m flux to be between 0.1 and 0.7 times that of P2, i.e. $1.05\pm0.8$\,Jy (it is listed as `P3' in Table \ref{table:sourcelist}). 

The MIPSGAL team recently released mosaics of the 70\,$\mu$m survey. We use the mosaics that overlap with our field of view to find one emission peak of roughly 7\,Jy, corresponding to the S5 and S6 mid-infrared sources (see Sect.~\ref{sec:stageIobjects}). There are no other 70~$\mu$m peaks, yielding a conservative upper limit of 5 Jy for the other sources in Table \ref{table:sourcelist}.

Ten sources have counterparts in one or all of the near-infrared J, H and K$_\mathrm{s}$ 2MASS bands \citep{skrutskie2006}, see Table \ref{table:sourcelist}. For the objects without 2MASS counterparts that show ambiguous model fitting results using only the Spitzer data points (S5, S10, S16 and S20), 2MASS upper limits (magnitude 15.8, 15.1 and 14.3 for J, H and K$_s$ bands, respectively) are included as constraining parameters.

\section{Results}
\label{sec:results}

\subsection{Fitting data points to models of Young Stellar Objects}
\label{sec:modelfitting}

The physical and chemical parameters and processes that play a role in specific YSO candidates are best explored by comparing the data points (flux values at various wavelengths) to models of YSOs. We use the models from \cite{robitaille2006}, who used a numerical Monte Carlo radiation transfer code \citep{whitney2003_II,whitney2003_I,whitney2004} to calculate a set of 20\,000 models for YSOs. In this grid of models, 14 physical parameters characterize the three main components of a YSO: (i) the central star, (ii) a rotationally flattened infalling envelope with bipolar cavities, and (iii) a flared accretion disk. The effective number of parameters is 8 for models that have an envelope but no disk, and 10 for models that have a disk but no envelope. Apart from the 14 parameters used as input for the radiative transfer code, 10 different viewing angles are considered, ranging from edge-on to pole-on with respect to the axisymmetric geometry of all modeled systems. Every viewing angle results in a different Spectral Energy Distribution (SED) of the same model; the grid therefore encompasses 10$\times$20\,000 different SEDs. Various star formation `modes' and stages, as well as environments, are taken into account by using a large spread in the input parameters. The code is developed bearing in mind the specific goal of applying the model grid to archives of \textit{Spitzer Space Telescope} data, such as those of GLIMPSE \citep{benjamin2003} and MIPSGAL \citep{carey2006}. This makes the model grid particularly useful for the interpretation of our data. 

An SED fitting tool \citep{robitaille2007} is available online\footnote{{\tt http://caravan.astro.wisc.edu/protostars/}}. Based on user-specified data points (flux and flux uncertainty in at least three passbands), the tool determines which of the models from the pre-computed grid \citep{robitaille2006} fits best to the data. Upper limits in certain bands may also be specified, but do not contribute to $N$, the number of data points. In addition, the user is asked to specify a fiducial distance range and a range in visual extinction, $A_\mathrm{V}$. 

For every model in the grid, the specified measured fluxes ($F_\nu(\lambda_i)$) are compared to convolved, scaled and extincted model fluxes ($P_\nu(\lambda_i)$), using $\chi^2$ as a measure for the deviation of a certain model from the specified data points:
\begin{equation}
\label{eq:fitchi2}
\frac{\chi^2}{N} = \frac{1}{N} \sum\limits_{i=1}^N \left( \frac{\langle \log_{10}(F_\nu(\lambda_i)) \rangle - \log_{10}(P_\nu(\lambda_i))} {\sigma( \langle \log_{10}(F_\nu(\lambda_i)) \rangle) } \right)^2.
\end{equation}
The $\sigma$ factor in the denominator of Eq.~(\ref{eq:fitchi2}) is the square root of the `unbiased' flux variance in $^{10}\log$ space, which is related to the relative flux uncertainty in each data point supplied by the user. The best fit is considered to be the SED that results in the minimum value of $\chi^2$ for the given set of data points: $(\chi^2/N)_\mathrm{best}$. An object can only be fitted to the model grid if $N\geq3$, not counting upper limits.

In addition to a comparison of a set of measured fluxes to the model YSO SEDs, the same data points are fitted to stellar photosphere SEDs with effective temperatures ranging from 2000 to \mbox{50\,000 K}. This feature is useful in assessing whether a particular object might be explained by a normal stellar photosphere instead of a YSO model.

\subsection{Spectral Energy Distribution analysis}
\label{sec:SEDanalysis}

A total of 17 of the 20 sources from Table \ref{table:sourcelist}, aided by sub-millimeter, MIPS 70~$\mu$m and 2MASS data, have $N \geq 3$ and can be analyzed using the models and the fitter from \citet{robitaille2006, robitaille2007} described in Sect.~\ref{sec:modelfitting}. Three sources, S1, S3 and S10, do not have enough data points to use the fitting routine. This analysis results in four suspected photospheres and 13 YSOs, summarized in Table \ref{table:allYSOs}. 

\begin{figure}[h] 
	\resizebox{0.495\hsize}{!}{\includegraphics{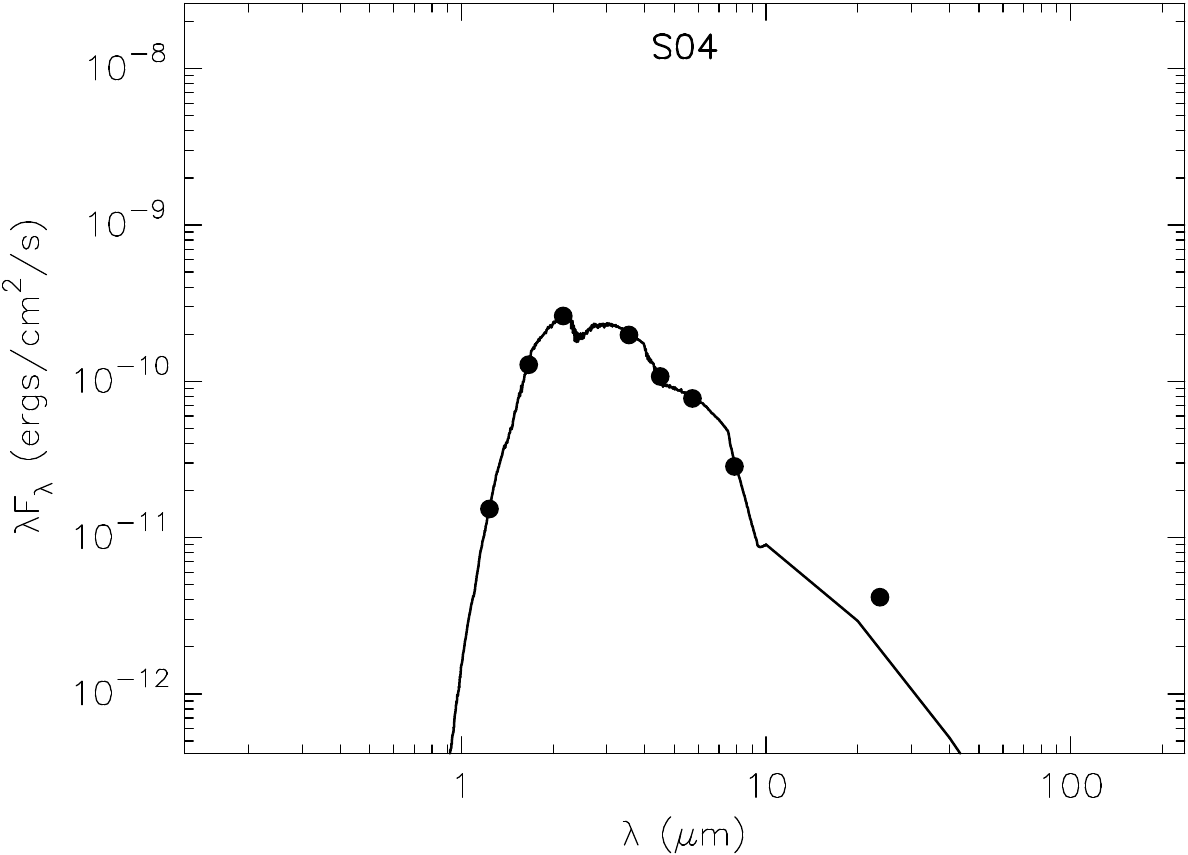}}
	\resizebox{0.495\hsize}{!}{\includegraphics{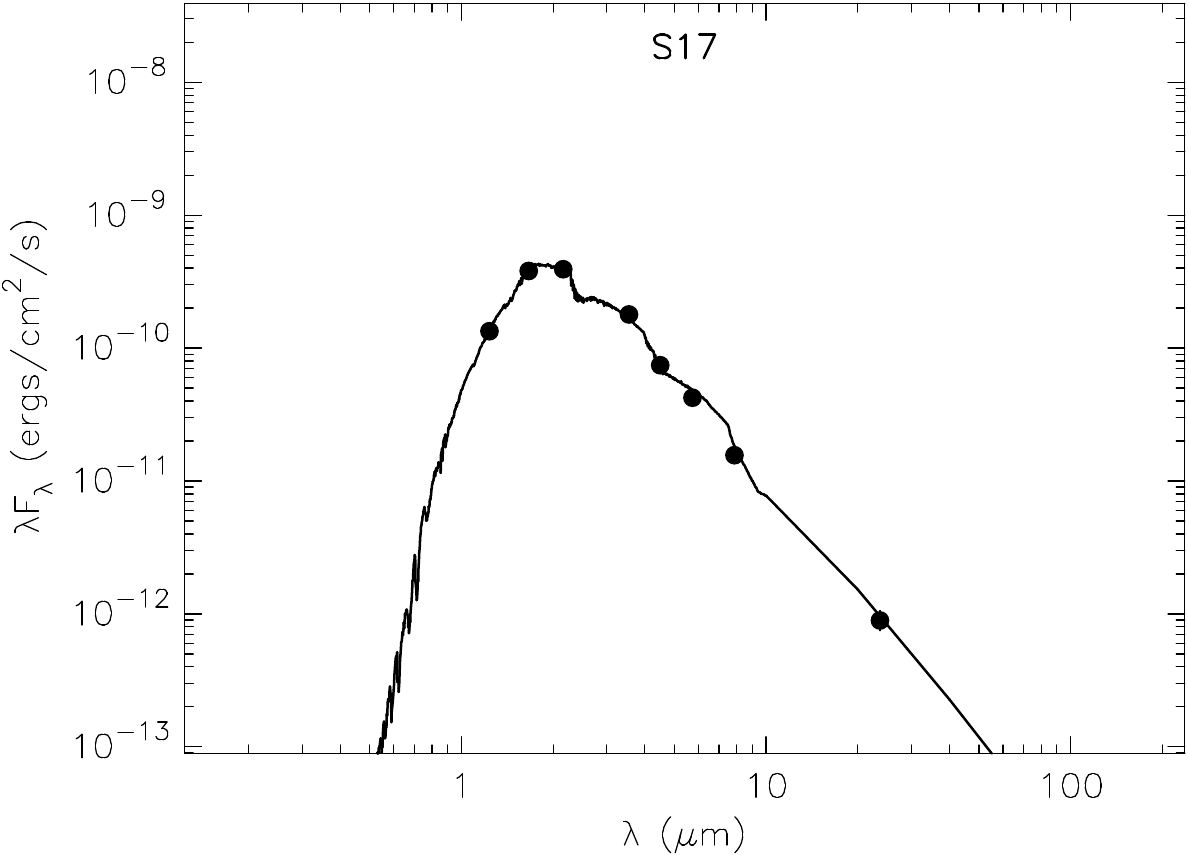}}
	\resizebox{0.495\hsize}{!}{\includegraphics{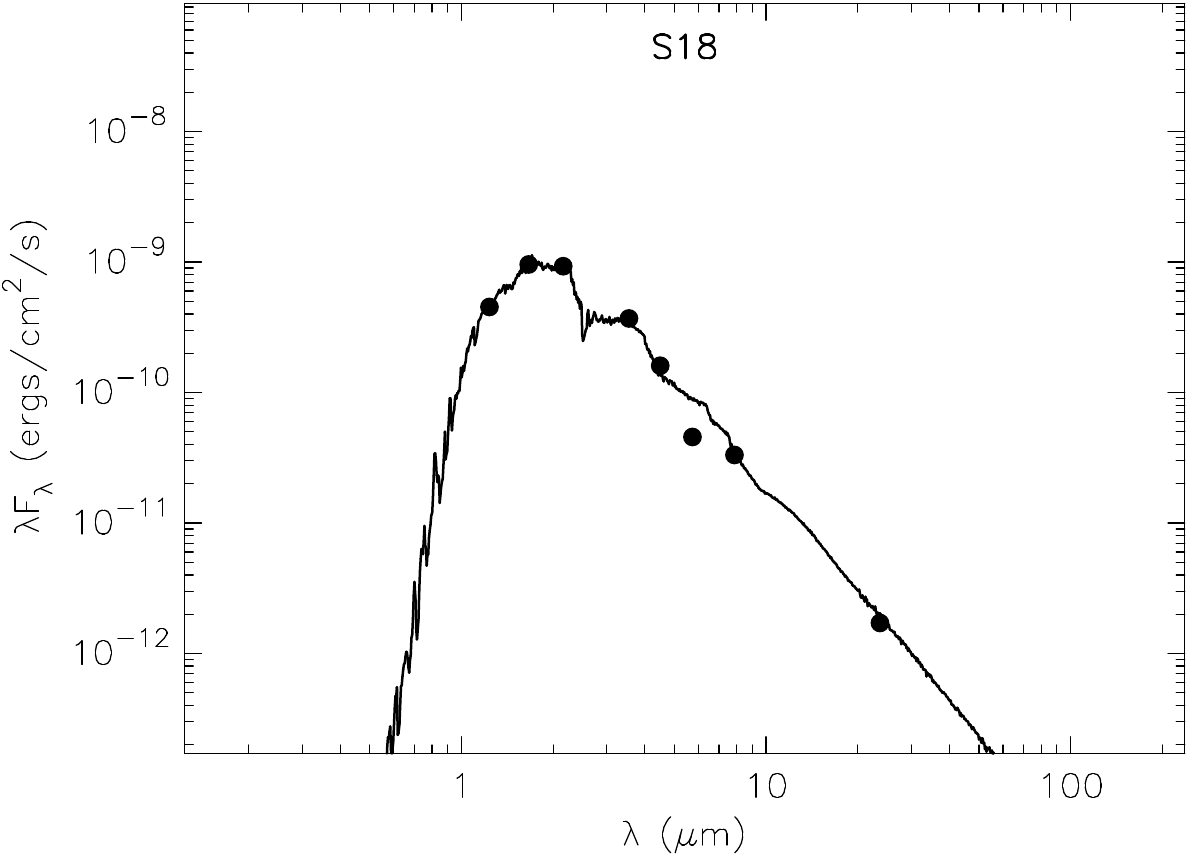}}
	\resizebox{0.495\hsize}{!}{\includegraphics{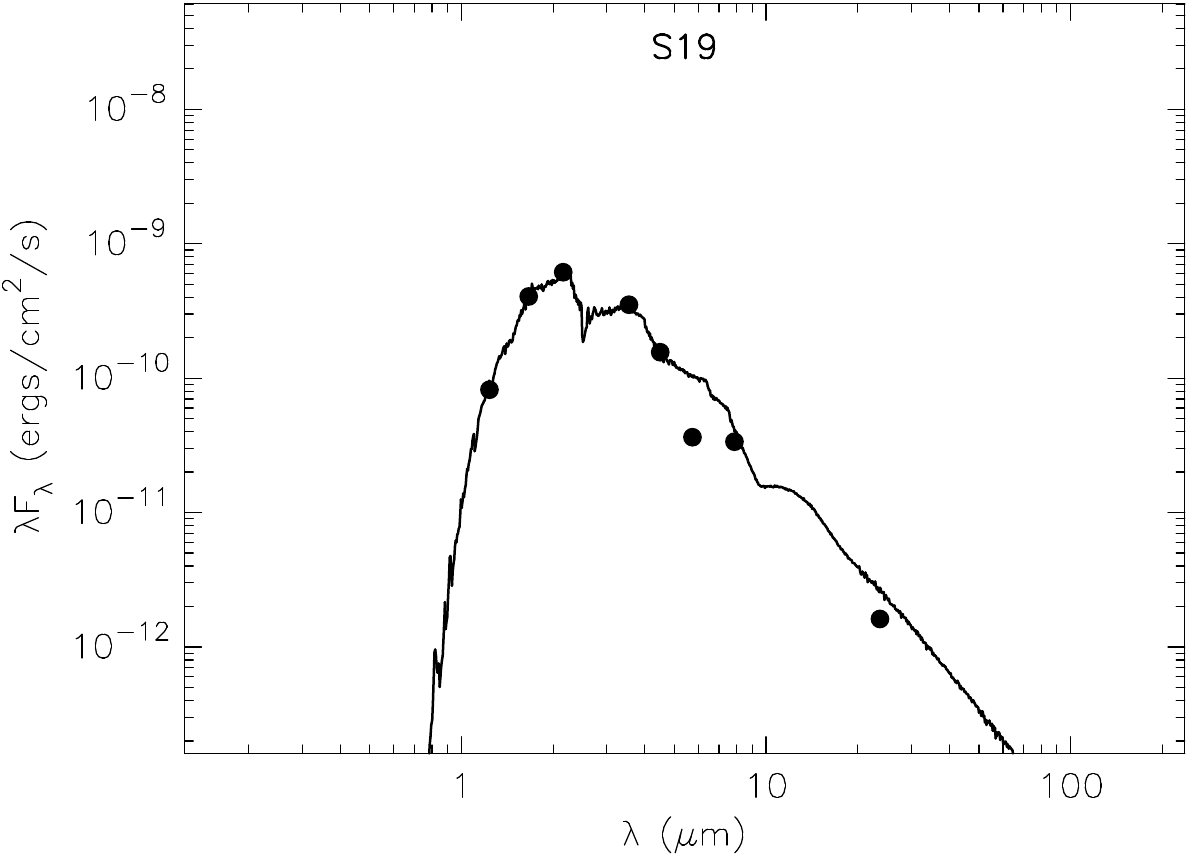}}
	\caption{Best fitting photosphere model (solid lines) to the data points (filled circles) of S4, S17, S18 and S19.} 
	\label{fig:photospherefits} 
\end{figure}

In order to separate YSOs from (extincted) photospheres, the data points of the 17 sources are compared first to photosphere models. Visual inspection of all SEDs identifies the objects S4, S17, S18 and S19 (SEDs are shown in Fig.~\ref{fig:photospherefits}) as photospheres. The $\chi^2$ value for the best fitting photosphere is considerably lower than for the best fitting YSO model for all four sources. These sources are suspected to be foreground sources along the line of sight toward G48.65. None of the photosphere objects seem to be associated with the filament of the IRDC in the plane of the sky (see Fig.~\ref{fig:MIPSsources}). Without any pre-conceived selection with respect to sky position, the other sources --~the YSO candidates~-- have a location on the sky closer to the dark cloud, where we expect star formation to occur. 

While AGB stars and even background galaxies are candidates to explain infrared excess in point sources, both types of objects are statistically unlikely to be present in the field of view considered in this paper. Using the model by \citet{wainscoat1992} for the spatial distribution of AGB stars in the Galactic disk, a crude volume integration along the pyramid subtended by the lines of sight toward the $5\arcmin \times 5\arcmin$ field yields an estimate of less than 1 AGB object in our field. A search through the NASA/IPAC extragalactic database\footnote{The NASA/IPAC Extragalactic Database (NED) is operated by the Jet Propulsion Laboratory, California Institute of Technology, under contract with the National Aeronautics and Space Administration.} results in two known galaxies within our field of view, neither of which coincide with any of the objects listed in Table \ref{table:sourcelist}. Considering this, we assume here that all objects in our field that show infrared excess are in fact YSOs.

The data points of the remaining 13 sources are compared to the YSO model grid of \citet{robitaille2006} (see Sect.~\ref{sec:modelfitting}). We attempt to derive an evolutionary stage for each source, based on three key parameters of the fitted models: the rate of mass accretion from the circumstellar envelope ($\dot{M}_\mathrm{env}$) and the mass of the circumstellar disk ($M_\mathrm{disk}$), both relative to the mass of the central object, $M_\star$. We prefer the classification of YSOs into evolutionary ``Stages" from \citet{robitaille2006} over the older ``Class 0/I/II/III" classification scheme, which is based solely on the slope of the mid-infrared SED \citep{adams1987,white2007}. The Stage classification provides a more physical basis for YSO classification: a source is assigned to a certain stage based on parameters of the model(s) that fit best to the data points. The threshold values for $\dot{M}_\mathrm{env}/M_\star$ and $M_\mathrm{disk}/M_\star$ are given in Table \ref{table:stages}. We choose to make no distinction between ``Stage 0" (with an SED resembling a 30 K graybody) and ``Stage I" (a protostar with a large accreting envelope), since the energy distributions of the two are difficult to distinguish in the mid-infrared; these two phases will be collectively referred to as ``Stage I" in the rest of this paper. A Stage I object is dominated by the accreting envelope; the envelope accretion has diminished significantly in the consecutive Stage II, which shows signs of a massive circumstellar disk emerging from the envelope; in Stage III, the disk has become less and less massive and the direct radiation from the central star begins to dominate the observable flux.

\begin{figure*}[!p]
\includegraphics[width=0.32\textwidth]{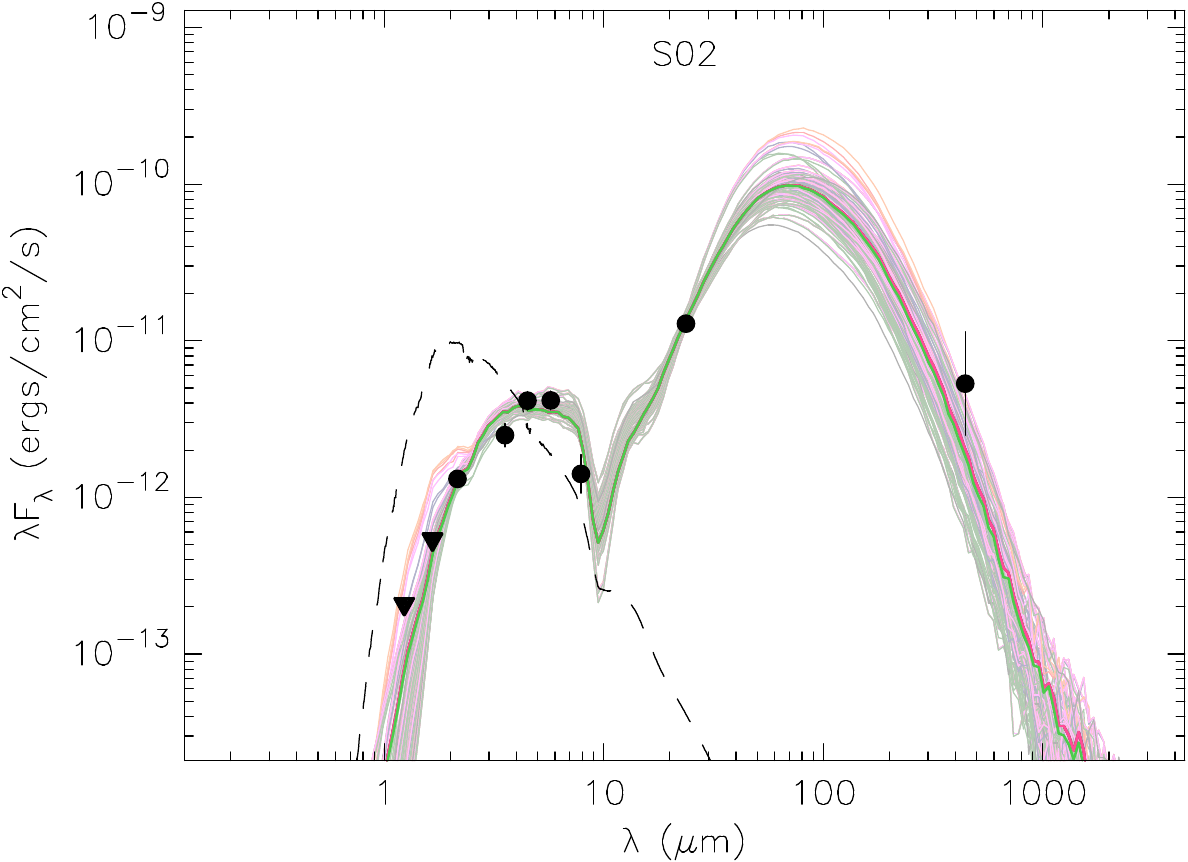}
\includegraphics[width=0.32\textwidth]{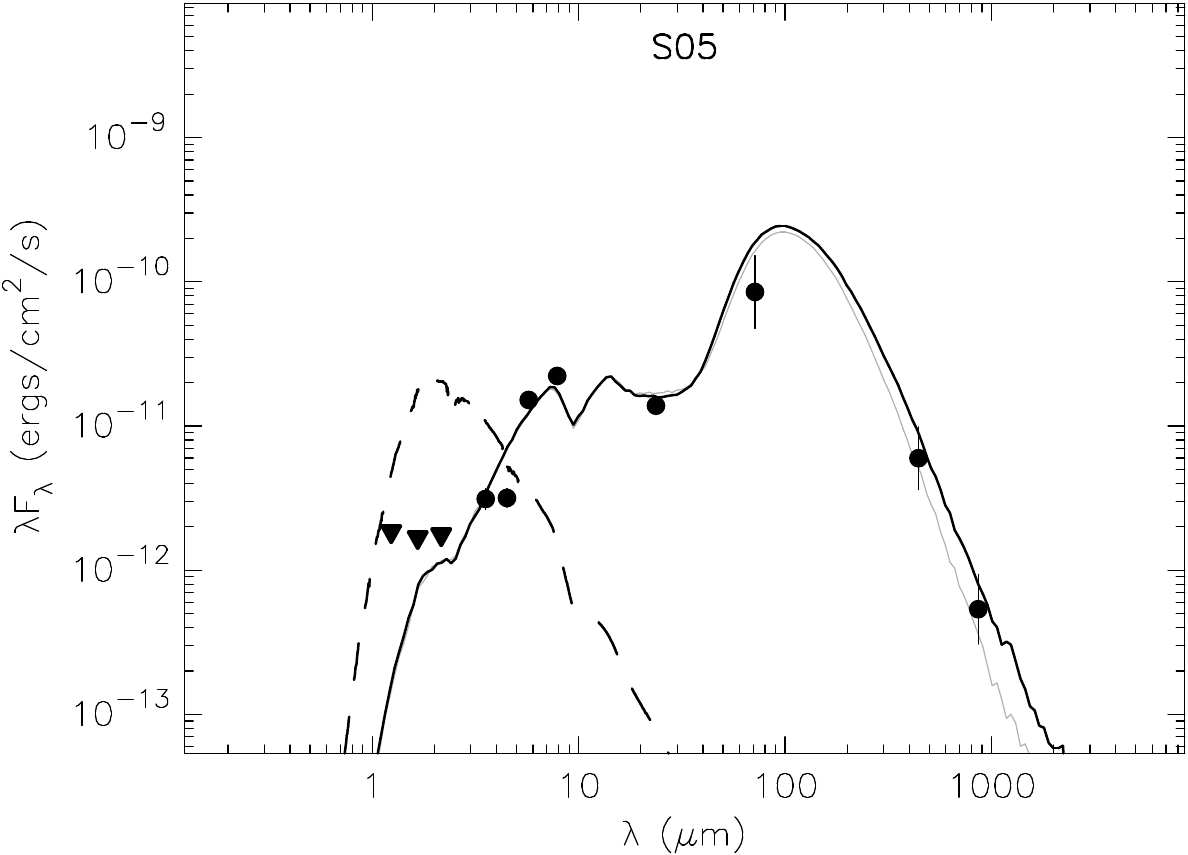}
\includegraphics[width=0.32\textwidth]{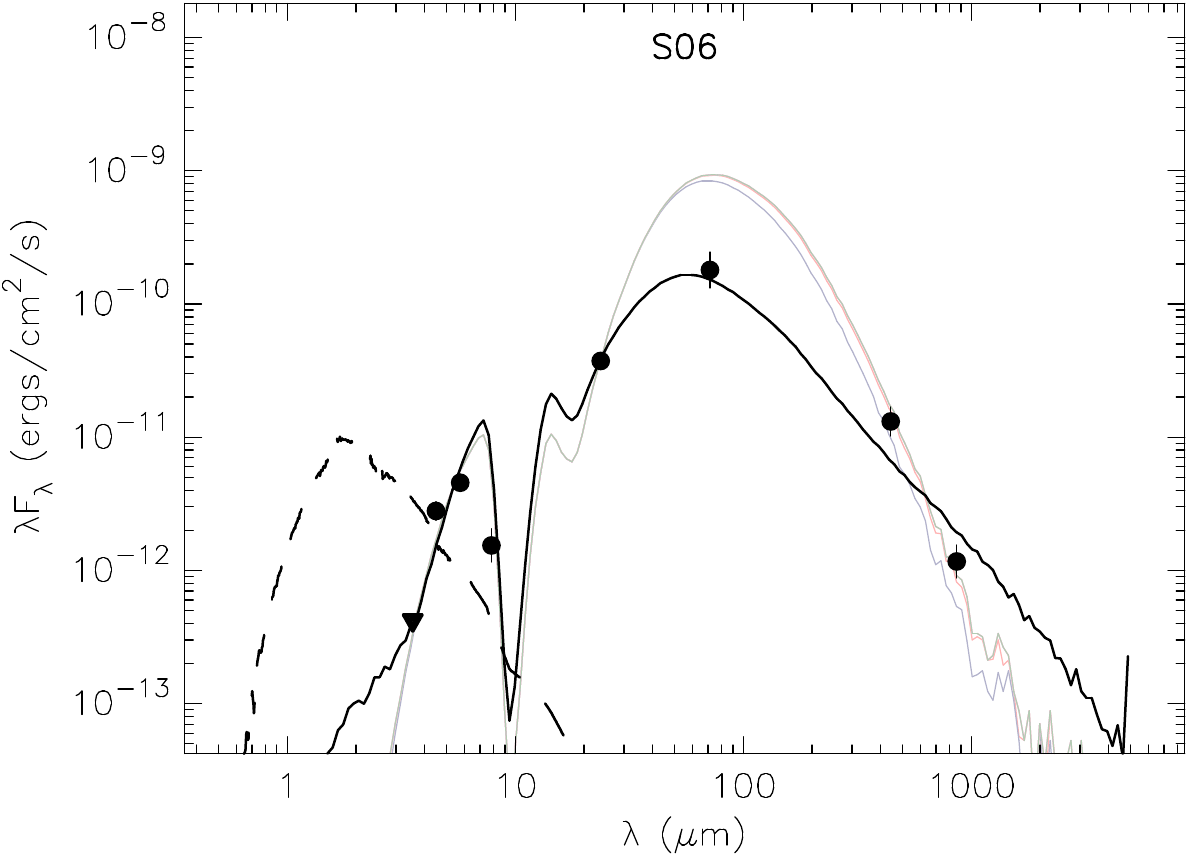}
\includegraphics[width=0.32\textwidth]{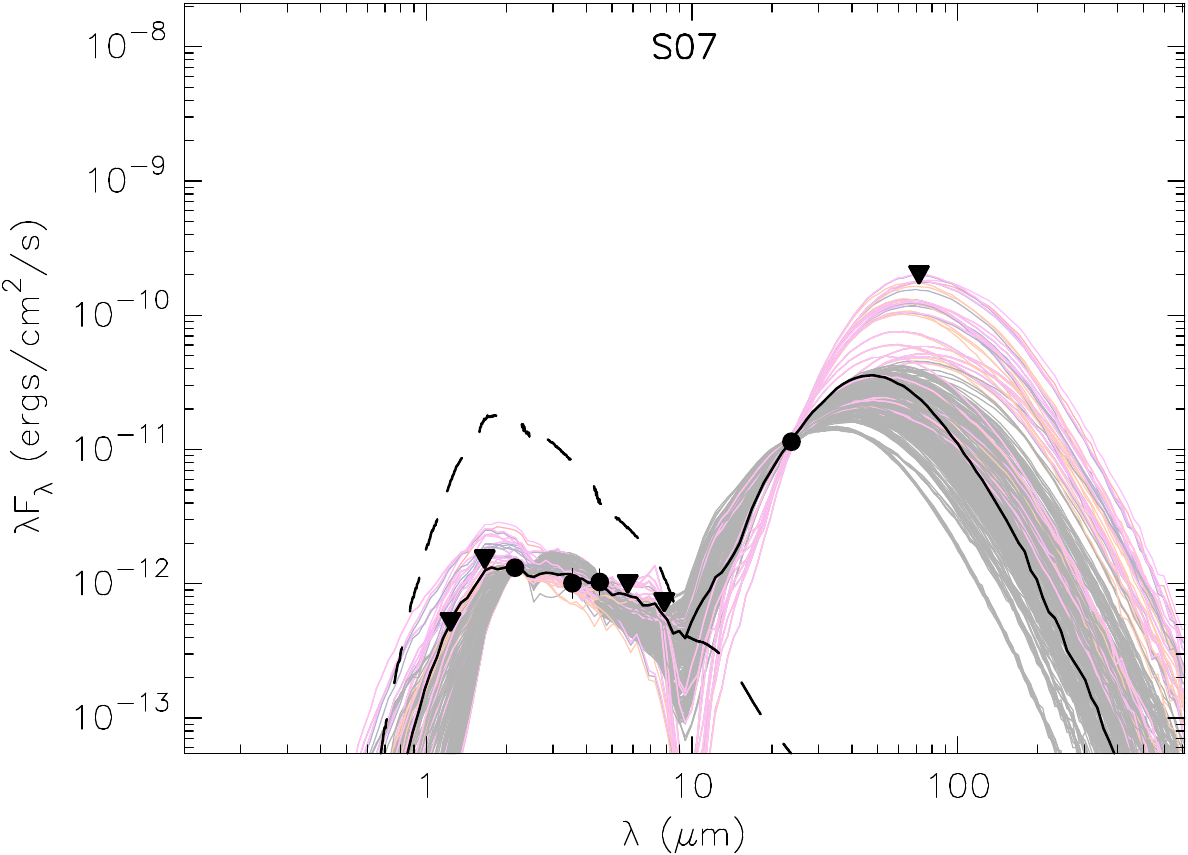}
\includegraphics[width=0.32\textwidth]{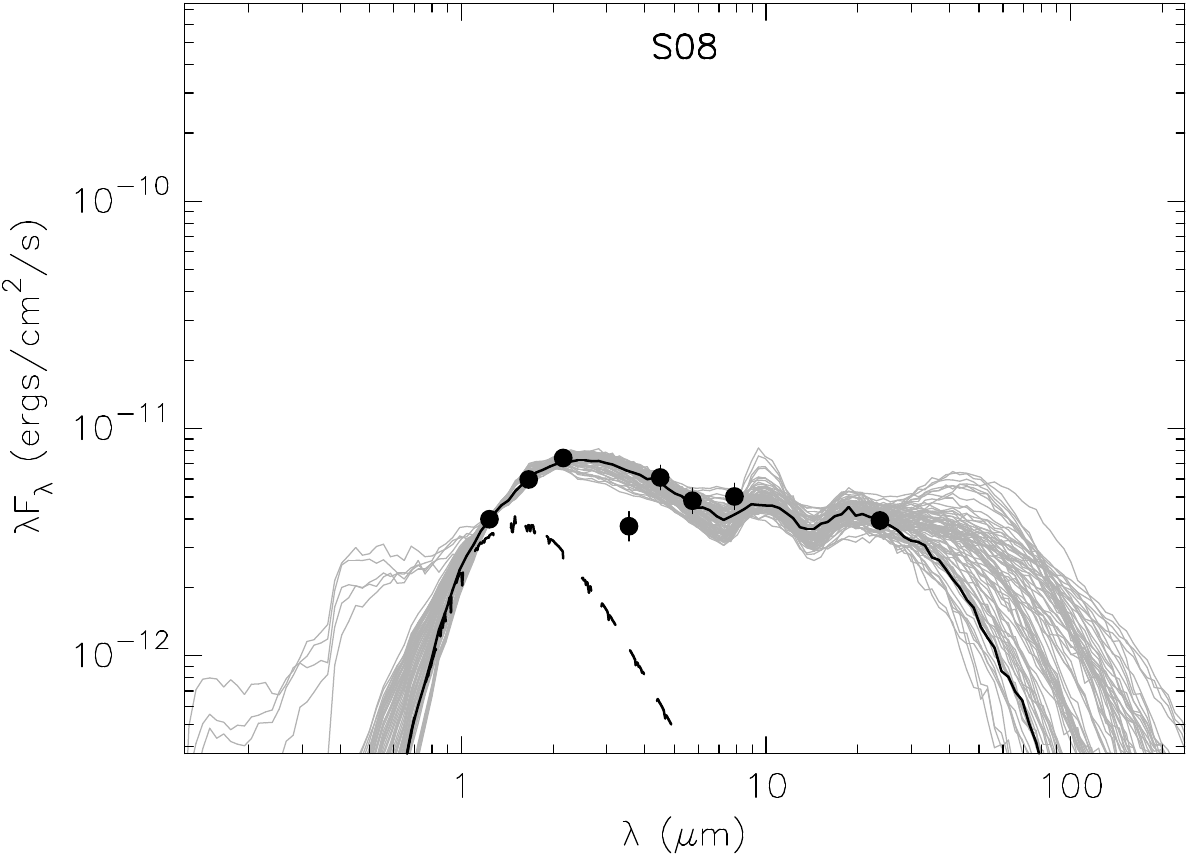}
\includegraphics[width=0.32\textwidth]{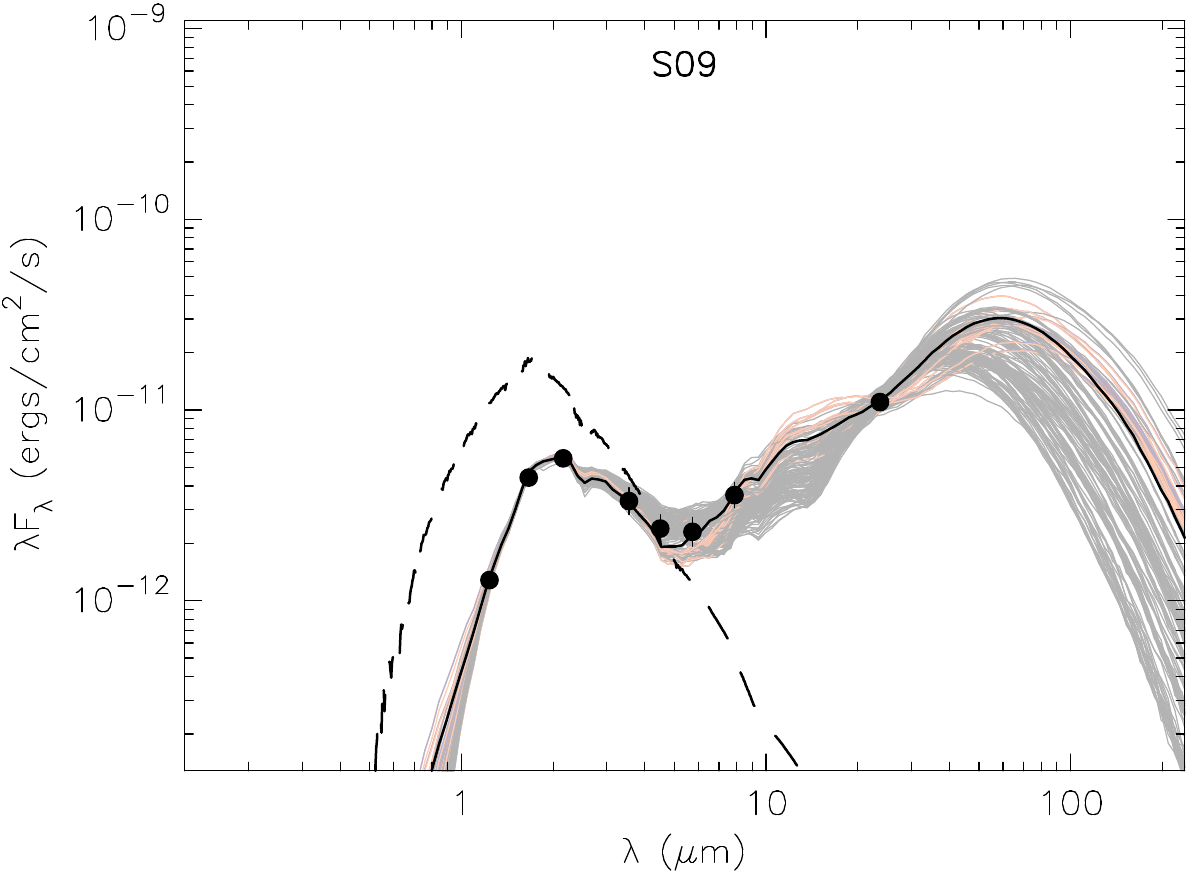}
\includegraphics[width=0.32\textwidth]{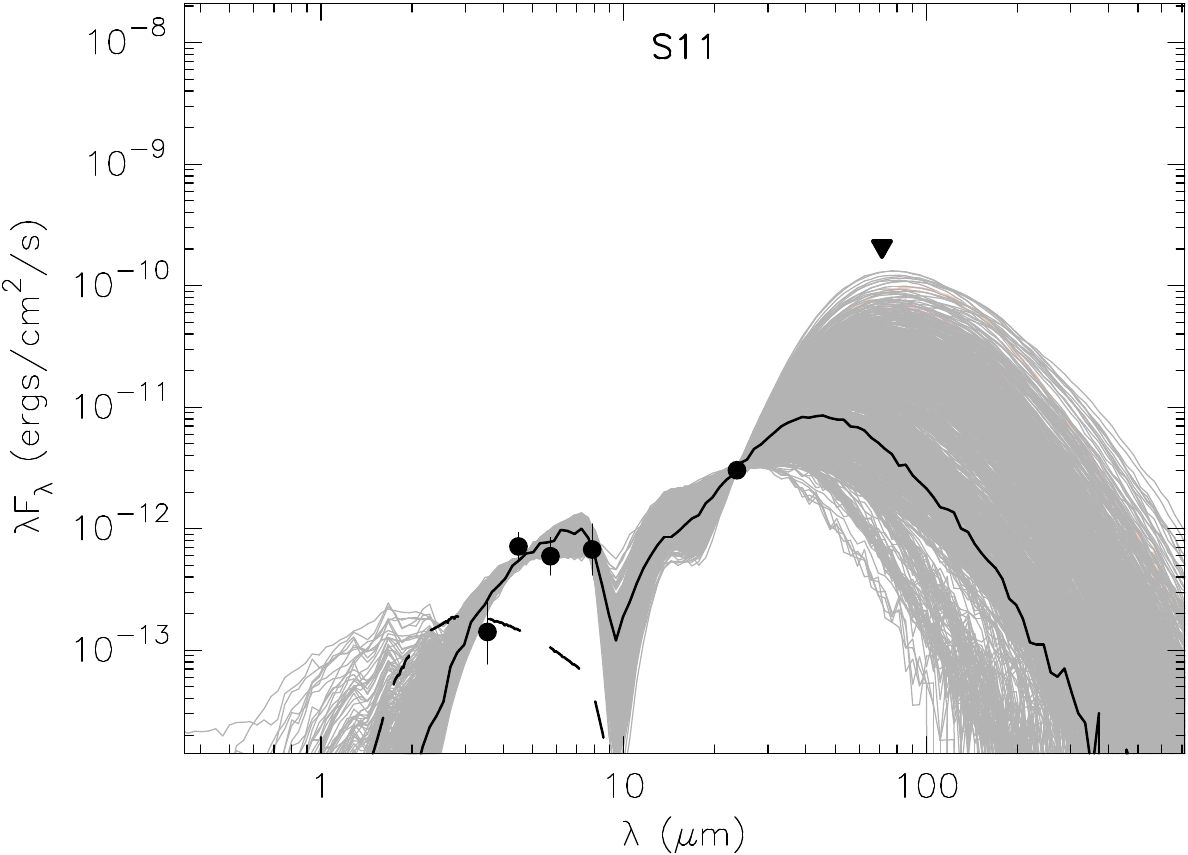}
\includegraphics[width=0.32\textwidth]{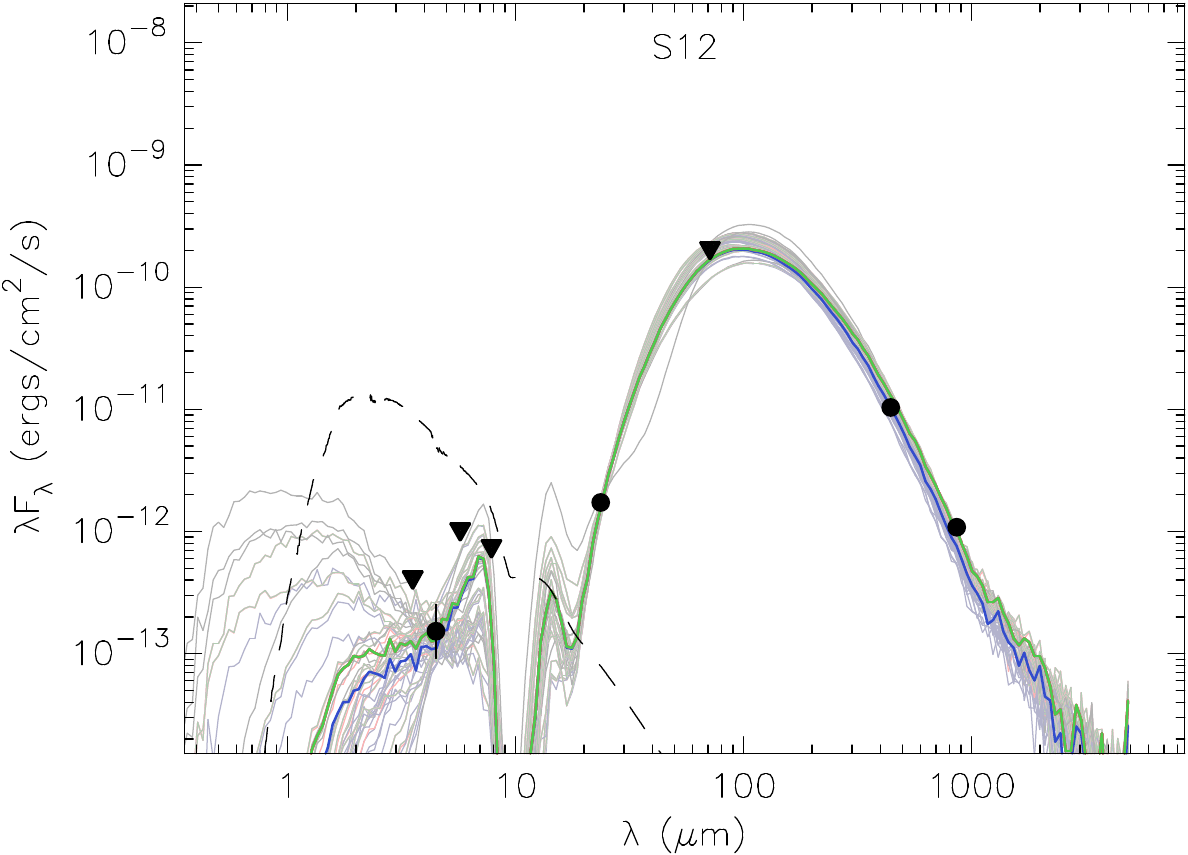}
\includegraphics[width=0.32\textwidth]{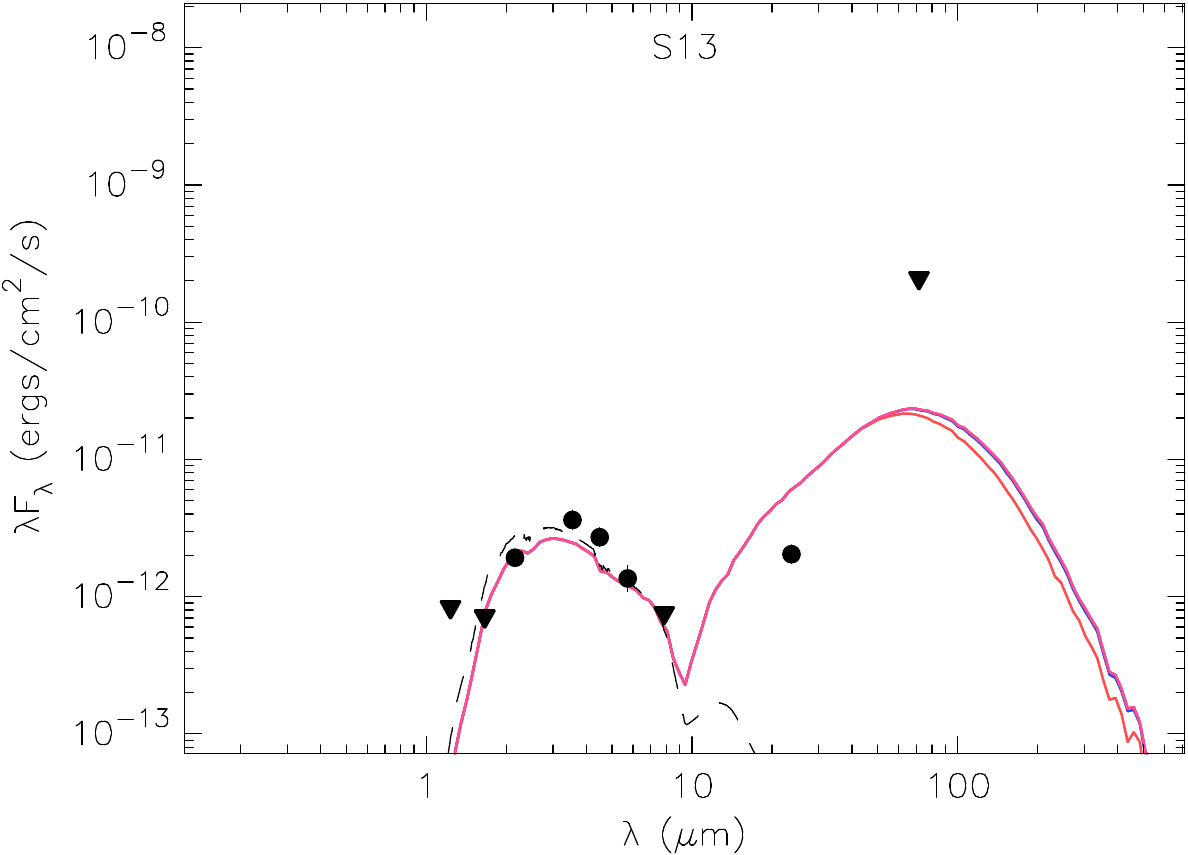}
\includegraphics[width=0.32\textwidth]{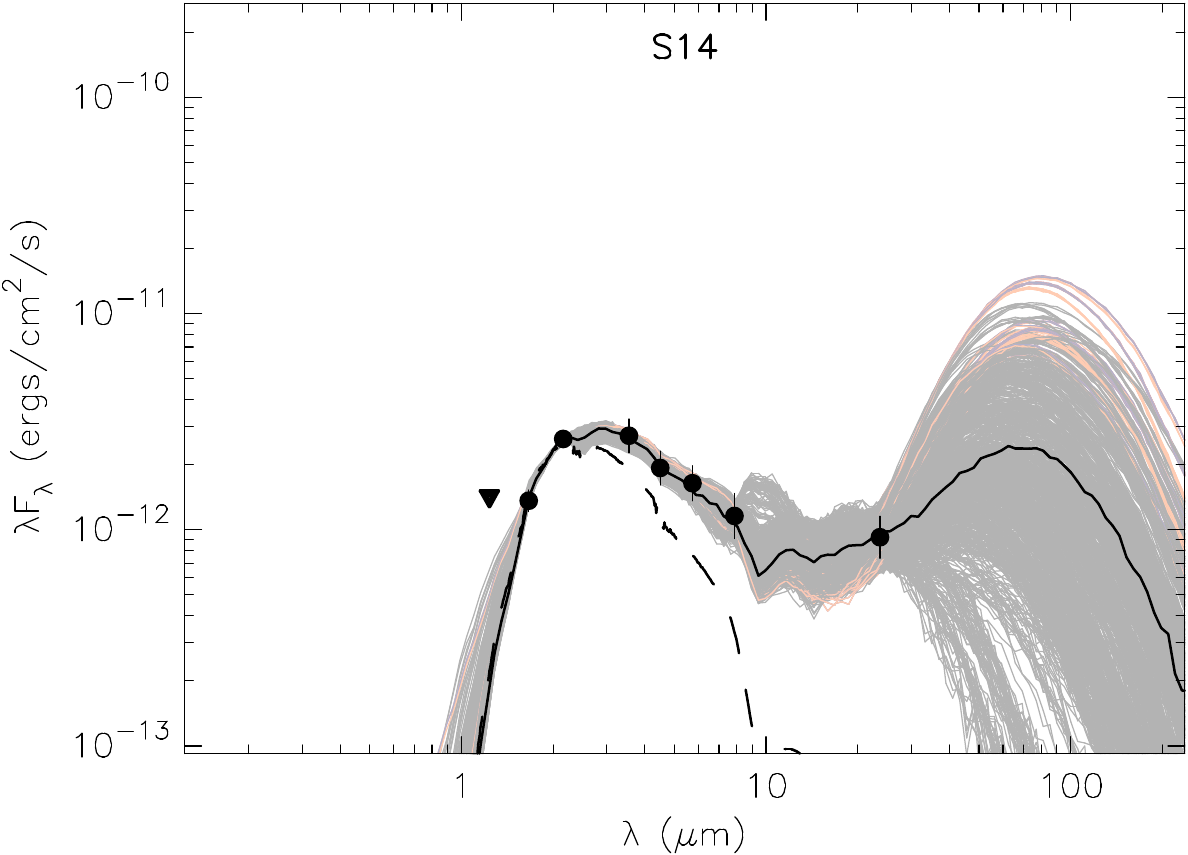}
\includegraphics[width=0.32\textwidth]{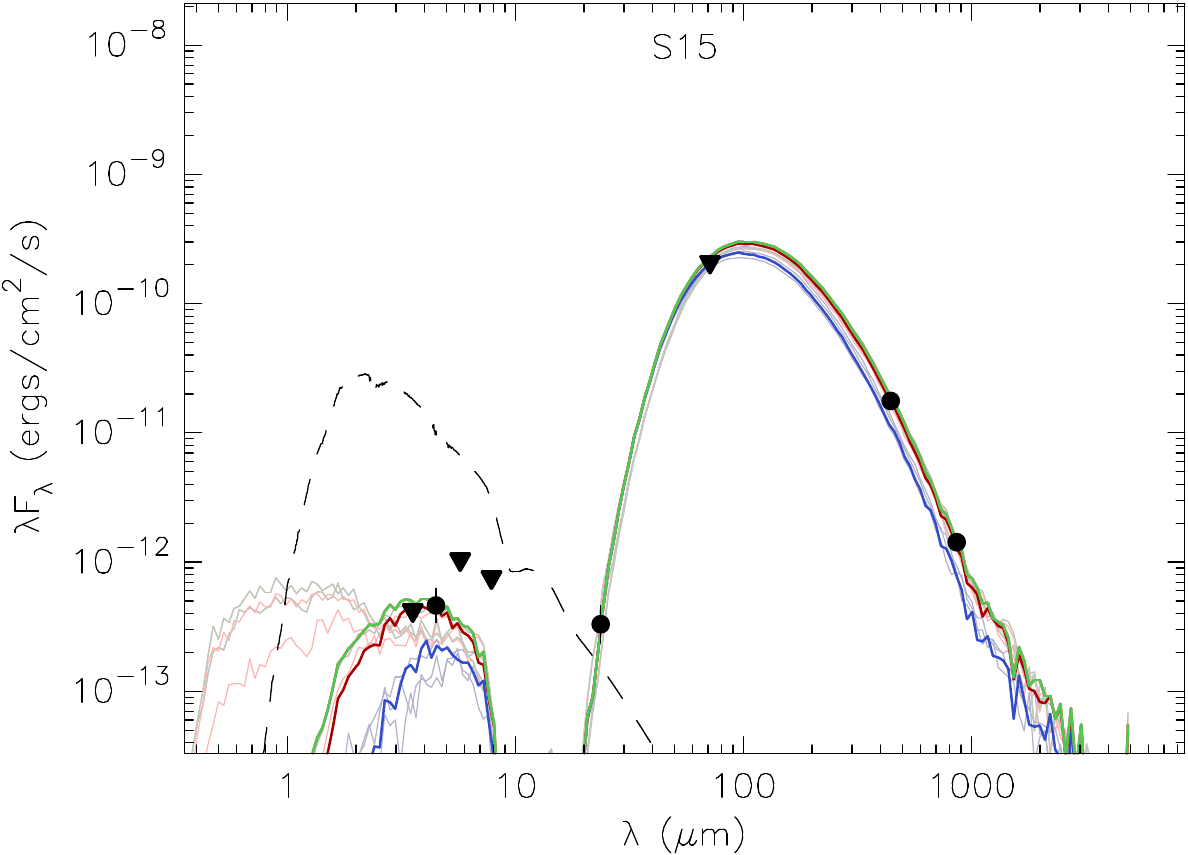}
\includegraphics[width=0.32\textwidth]{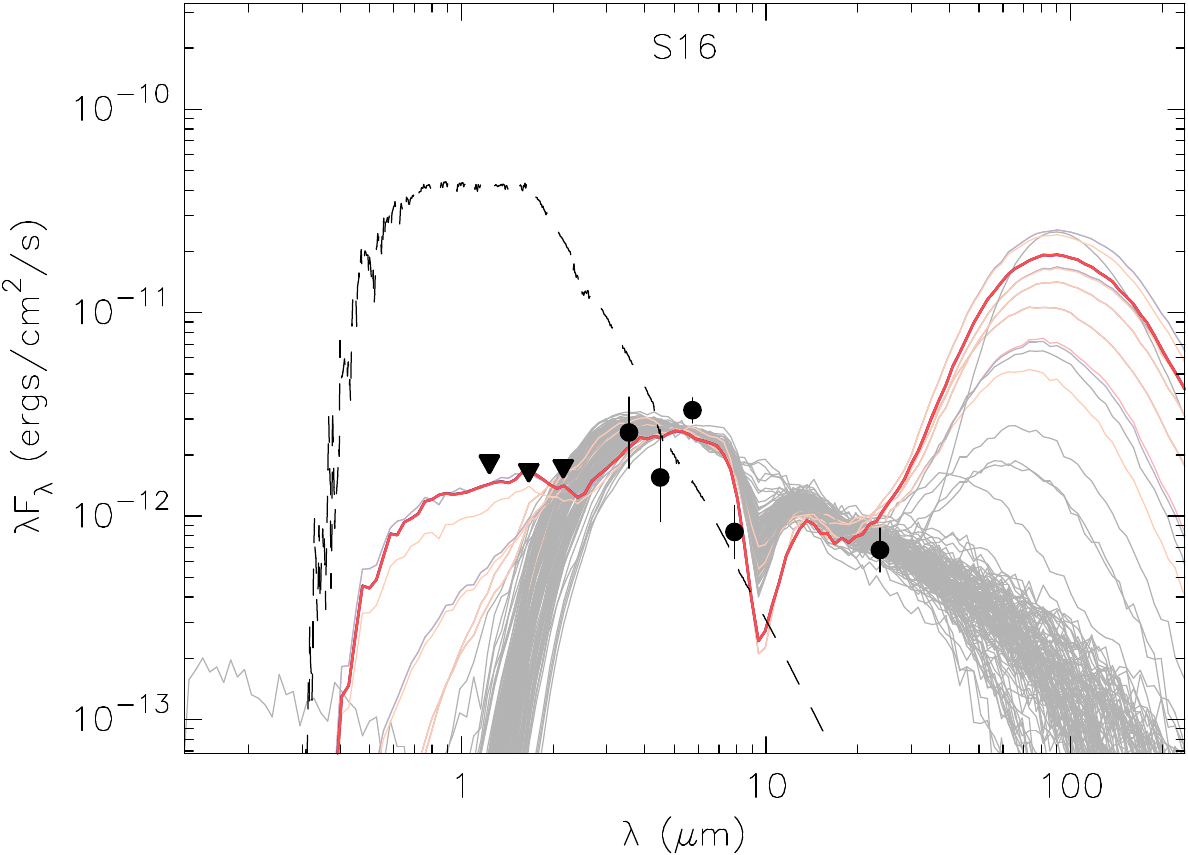}
\includegraphics[width=0.32\textwidth]{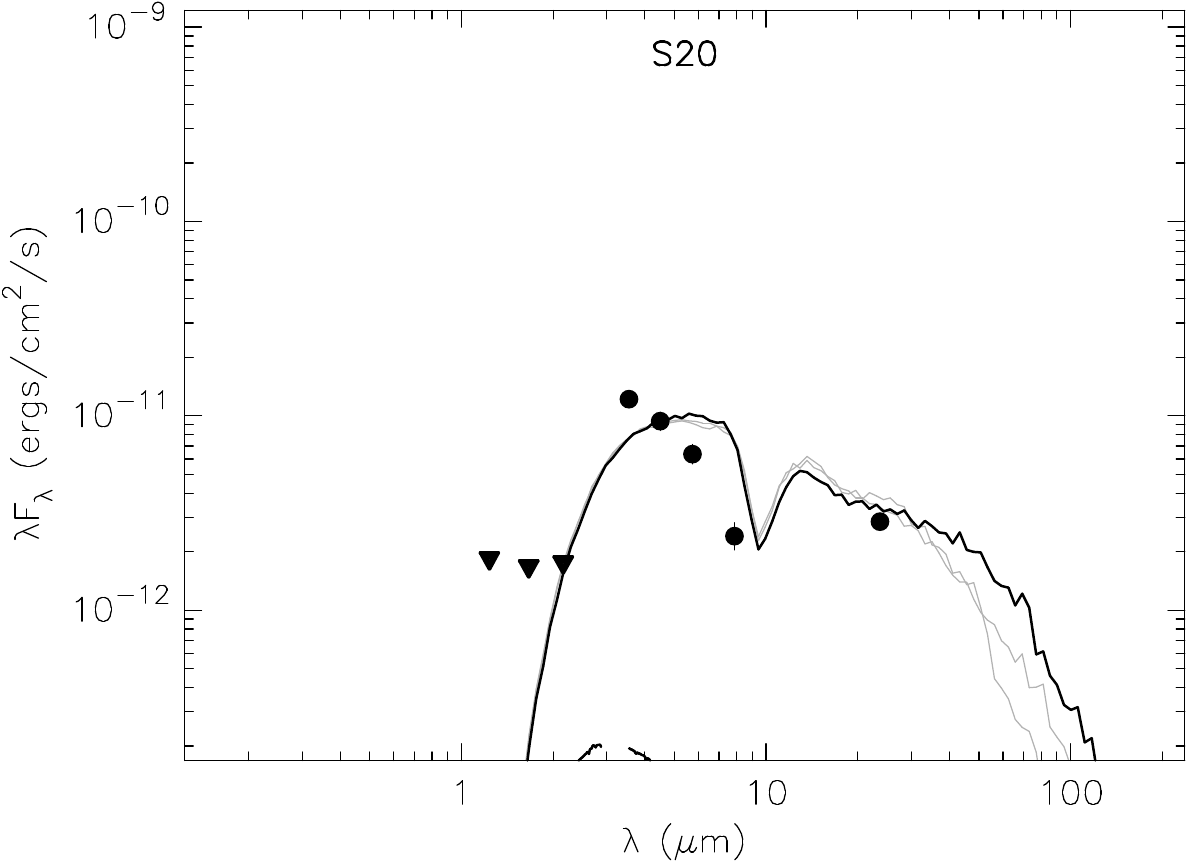}
\hspace{8.95cm}
\includegraphics[width=0.18\textwidth]{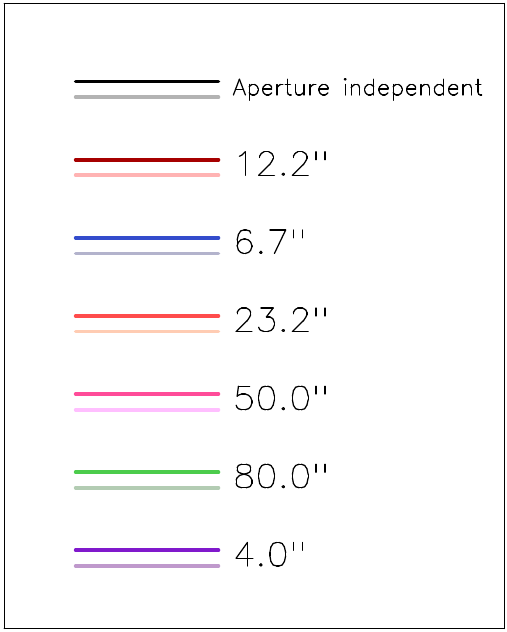} 
\caption{YSO model SEDs that fit best to the data points (filled circles) of thirteen sources. This figure shows those sources listed in Table \ref{table:allYSOs}. Objects S2, S5, S6, S7, S9 and S12 are discussed in detail in Sect.~\ref{sec:stageIobjects}, objects S8, S11, S14 and S16 in Sect.~\ref{sec:stageIIobjects} and objects S13 and S20 in Sect.~\ref{sec:uncertainobjects}. In each panel, the black or colored curve is the best fitting SED (if it is colored, this indicates dependence on aperture size, see legend in bottom right frame). The lighter shaded curves (gray or pale colors) are all other model SEDs for which the criterion in Eq.~(\ref{eq:chi2criterion}) is obeyed. The dashed curve is the photosphere used as input for the radiative transfer code that calculated the best fitting YSO SED.}
	\label{fig:YSOSEDs}
\end{figure*}

\begin{table*} 
\begin{minipage}[t]{\textwidth}
\caption{Sources classified as YSO.}
\label{table:allYSOs}
\centering
\renewcommand{\footnoterule}{}  
\begin{tabular}{r l l l l l l l}
    \hline\hline
   (1) & (2) & (3) & (4) & (5) & (6) & (7) & (8)\\
   ID & assoc. & $N$ & minimal $\chi^2/N$ & $M_\star$ & $\dot{M}_\mathrm{env}/ M_\star$ & $M_\mathrm{disk}/M_\star$ & Stage \\
  &  	&	&	& 	& ($M_\odot$)	& (yr$^{-1}$) &  \\
    \hline
    S2  & yes   & 7 & 1.38  & 1--5  &  $10^{-5}$--$2\times10^{-4}$  &  $3\times10^{-5}$--$7\times10^{-2}$   & I\\
    S5  & yes   & 8 & 6.18  & 1--2 & $1$--$3\times10^{-4}$  & $1$--$5\times10^{-2}$ & I \\
    S6  & yes   & 7 & 6.77  & 0.7--8  & $1$--$6\times10^{-5}$  & $3\times10^{-6}$--$5\times10^{-2}$ & I \\
    S7  & yes   & 4 & 0.0875 & 0.2--6 & $5\times10^{-7}$--$2\times10^{-4}$ & $8\times10^{-5}$--$6\times10^{-3}$ & I\\
    S8  & yes   & 8 & 1.99  & 1.5--4 & 0--$5\times10^{-8}$     &  $ 5\times10^{-6}$--$4\times10^{-2}$ & II\\
    S9  & yes   & 8 & 0.126 & 0.3--6 & $2\times10^{-7}$--$3\times10^{-4}$ & $5\times10^{-6}$--$7\times10^{-2}$ & I or (II) \\
    S11 & yes   & 5 & 0.61 & 0.1--5 & 0--$2\times10^{-4}$ & $5\times10^{-8}$--$2\times10^{-1}$ & I or II or (III) \\
    S12 & yes   & 4 & 0.58 & 1.5--6 & $1$--$4\times10^{-4}$ & $2\times10^{-4}$--$3\times10^{-2}$ & I  \\
    S13 & yes   & 5 & 11.9 & $\sim$2  & $\sim10^{-6}$  & $\sim$$10^{-2}$  & $\sim$I \\
    S14 & yes   & 7 & 0.030 & 0.5--4  & 0--$4\times10^{-5}$  & $3\times10^{-8}$--$5\times10^{-2}$ & (I) or II or (III)\\
    S15 & yes   & 4 & 0.515 & 3--4 & $2\times10^{-4}$ & $3\times10^{-4}$--$5\times10^{-3}$ & I \\
    S16 & yes   & 5 & 1.99 & 2--5 & 0--$3\times10^{-4}$ & $3\times10^{-6}$--$4\times10^{-2}$ & I or II \\
    S20 & no    & 5 & 18.6 & $\sim$4--6 & $\sim$0 & $\sim$$10^{-5}$--$10^{-3}$ & $\sim$II \\
    \hline
\end{tabular}
\begin{list}{}{}
	\item 
	Object parameters derived from fitting the data points to YSO SEDs (see Sec.~\ref{sec:SEDanalysis}). Column (1) lists the ID of the object, corresponding to the entry in Table \ref{table:sourcelist}. Column (2) indicates whether the source is thought to be spatially associated with the IRDC filament. The number of available data points, $N$, is listed in column (3). The $\chi^2$ per data point, $\chi^2/N$, is given in column (4). Columns (5), (6) and (7) list the fitted stellar mass, envelope accretion rate and disk mass, respectively. Note that the latter two are given in terms of $M_\star$, not $M_\odot$. Finally, column (8) lists the Stage in which the sources fall according to the values in columns (6) and (7) and the Stage definition in Table \ref{table:stages}. If multiple possibilities are given for the Stage of a source, the less likely one(s) is (are) presented in brackets. A `$\sim$' prefix in column (5), (6), (7) or (8) indicates a high uncertainty in the derived parameter, resulting either from a high value of the minimal $\chi^2/N$ and/or a large spread in fitted parameter values. 
\end{list}
\end{minipage}
\end{table*}

Table \ref{table:allYSOs} lists the 13 objects that are classified as YSOs. The spread in model parameters $M_\star$, $\dot{M}_\mathrm{env}$ and $M_\mathrm{disk}$ given in the table is obtained by considering all models that obey the criterion:
\begin{equation}
\label{eq:chi2criterion}
\frac{\chi^2}{N} < \left( \frac{\chi^2}{N} \right)_\mathrm{best} + 1,
\end{equation}
with $\chi^2/N$ as defined in Eq. (\ref{eq:fitchi2}). This criterion is chosen after visual inspection of the sets of models that fit the data points of each of the objects: it should not be so stringent that only one model fits a source, and on the other hand, it should not be so indiscriminate that it allows the majority of the models in the grid to be a ``good fit". It is subsequently kept fixed for all sources. 
If Table \ref{table:allYSOs} lists only one value for a certain parameter, the value is implied to be known to an accuracy indicated by the number of significant digits. We refrain from defining a proper confidence level for each derived parameter value, since the 14 dimensional space of physical parameters is too sparsely sampled to derive any formal confidence intervals \citep{robitaille2007}. The objects classified as YSOs are discussed in detail below, starting with the Stage I objects. 

\begin{table} 
\caption{YSO evolutionary stages as defined by \citet{robitaille2006}.}
\label{table:stages}
\centering
\begin{tabular}{l l r r}
    \hline
    \hline
    Stage   & components                & $M_\mathrm{disk}/M_\star$    & $\dot{M}_\mathrm{env}/M_\star$ \\
    \hline
    I    & accreting envelope        & any                       & $>10^{-6} \ \mathrm{yr}^{-1}$ \\
    II      & disk, remains of envelope & $> 10^{-6}$       & $< 10^{-6} \ \mathrm{yr}^{-1}$ \\
    III     & optically thin disk       & $< 10^{-6}$       & $< 10^{-6} \ \mathrm{yr}^{-1}$\\
    \hline
\end{tabular}
\end{table}

\subsubsection{Stage I: S2, S5, S6, S7, S9, S12 and S15}
\label{sec:stageIobjects}
The sub-millimeter core ``P1", identified by \citet{ormel2005}, is separated into two distinct sources by MIPS and IRAC (see Fig.~\ref{fig:G48_3color}). What \citet{ormel2005} modeled as one emission core, now turns out to be composed of at least two individual YSOs: S6 and S5 in this paper. The 24~$\mu$m peaks of these sources are separated by roughly 12\arcsec, about 0.15 pc tangential distance at 2.5 kpc distance. It is crudely assumed that $\sim$50--80\% of the 450 and $\sim$50--90\% of the 850 $\mu$m SCUBA flux of P1 originates from S6, the brightest of the two cores in 24~$\mu$m; the remainder of the P1 sub-millimeter flux is attributed to S5, with the same uncertainty margins. The large uncertainties are adopted mainly to reflect the uncertainty in the assumption of the relative contributions of S5 and S6 to the total P1 flux. In addition, the MIPSGAL 70 $\mu$m mosaic image \citep{carey2006} that covers G48.65 is found to contain an emission peak at the position of P1 (S5 and S6). Its flux is measured using the same method as for the IRAC and MIPS 24~$\mu$m bands (see Sect.~\ref{sec:PSE}). Its flux is found to be 6.9$\pm$0.3 Jy at 70~$\mu$m. Like for the sub-millimeter flux described above, it is uncertain how much of this flux is contributed by each individual mid-infrared source (S5 and S6 fall inside one MIPS 70~$\mu$m beam). The 70~$\mu$m flux is therefore attributed to S5 and S6 in the same proportion as the 450~$\mu$m flux, i.e. $\sim$50--80\% to S6 and $\sim$20-50\% to S5.

\begin{figure}  
    \resizebox{\hsize}{!}{\includegraphics{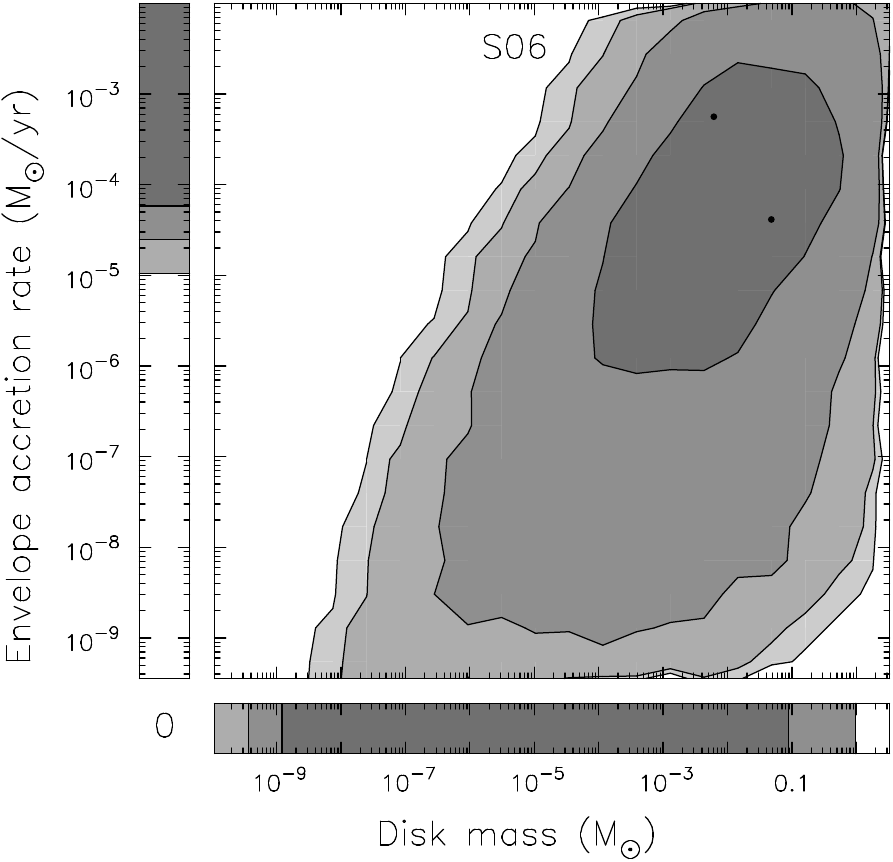}}
    \caption{Ranges of $\dot{M}_\mathrm{env}$ (vertical axis) and $M_\mathrm{disk}$ (horizontal axis) of models that fit the data of S6. The gray levels show the distribution of the models in the grid, where darker color indicates a higher density of models. The points show the loci of all models which obey the criterion in Eq.~(\ref{eq:chi2criterion}), i.e. the model SEDs which are plotted in Fig.~\ref{fig:YSOSEDs}.}
    \label{fig:S06_parameters}  
\end{figure}

Both S6 and S5 (see Fig.~\ref{fig:YSOSEDs} for the SED fits) are fit by models with high envelope accretion rates. Figure~\ref{fig:S06_parameters} shows the spread in $\dot{M}_\mathrm{env}$-$M_\mathrm{disk}$ space of models fitted to S6; the diagram looks similar for S5. Both sources are therefore classified as Stage I objects. The total luminosity according to the best-fitting models is $\sim$$10^2$--$10^3\,L_\odot$ for S6 and $\sim$$10^2\,L_\odot$ for S5. The sum is consistent with the 1-$\sigma$ limit of the total luminosity that was modeled for P1 by \citet{ormel2005}: $330^{+370}_{180}\,L_\odot$. 

The sources S15 \citep[sub-millimeter peak EP in][]{ormel2005} and S12 (P2) are only detected in MIPS 24~$\mu$m, in the sub-millimeter, very faintly in the IRAC~4.5$\mu$m band, and not in the other IRAC bands. Like S5 and S6, they are well fit by models with high values of $\dot{M}_\mathrm{env}$, exceeding $10^{-4}\ M_\odot/\mathrm{yr}$. Figure \ref{fig:YSOSEDs} shows the SED fits for S15 and S12. The inferred stellar masses for both objects are between 1 and 6 $M_\odot$, the total luminosities are $\sim$10$^2$ $L_\odot$ each. Although S12 (EP) appeared slightly less luminous from sub-millimeter continuum modeling, the rates of energy production inferred here for S12 and S15 are largely consistent with the modeled luminosity values for EP and P2 in Table 3 of \citet{ormel2005}, $300^{+1000}_{-230}$ and $19^{+110}_{-17} L_\odot$, respectively.

Model fitting for the S2 data points, including the 450~$\mu$m data point (see Sect.~\ref{sec:SCUBAand2MASS}), classifies it as a 1--5 $M_\odot$ Stage I object, see Table \ref{table:allYSOs}. S7 and S9, on the northern side of the cloud filament, are also classified as a Stage I objects in Table \ref{table:allYSOs}. A minority of the well fitting models lie in the Stage II region for both these objects. Nevertheless, it is deemed more likely that they are in Stage I.

Summarizing, all mid-infrared objects that show counterparts at 850~$\mu$m and/or 450~$\mu$m represent the very earliest stage of star formation. However, S7 and S9 -- with no data points longward of 30~$\mu$m -- show that a sub-millimeter counterpart is not a required condition for a source to be classified as Stage I.

\subsubsection{Stage II: S8, S11, S14 and S16}
\label{sec:stageIIobjects}

\begin{figure}  
    \resizebox{\hsize}{!}{\includegraphics{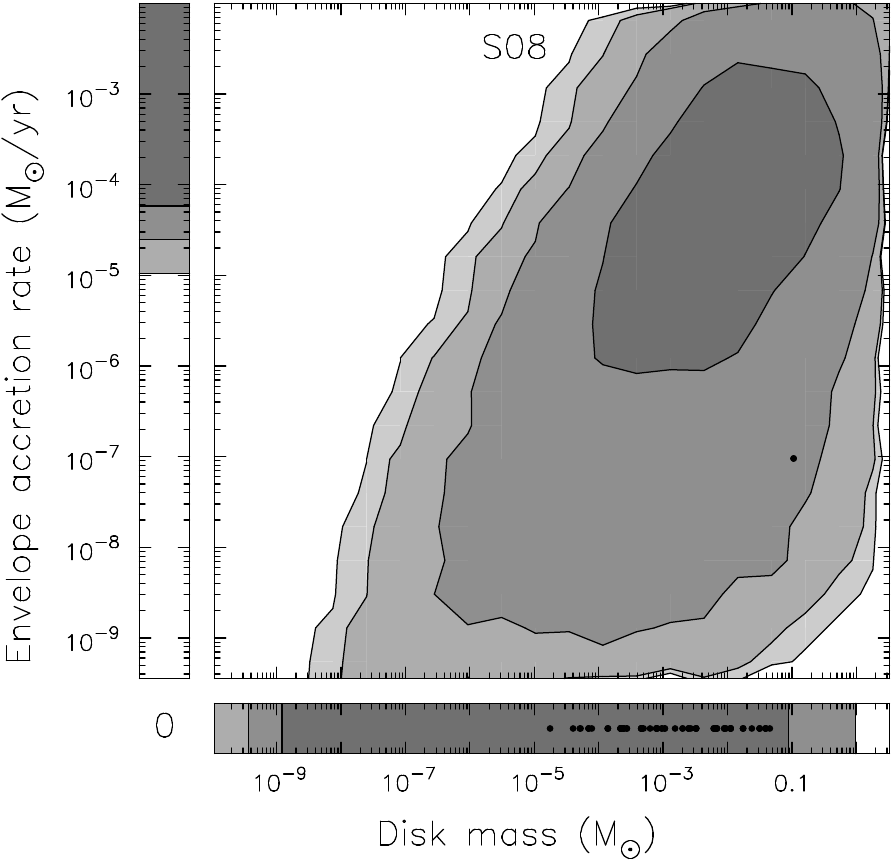}}
    \caption{Same as Fig.~\ref{fig:S06_parameters}, for S8.}
    \label{fig:S08_parameters}  
\end{figure}

S8 is found to match models with a stellar mass of 1.5--4 $M_\odot$, and no accretion from the envelope. The absence of envelope accretion, shown by the distribution in $\dot{M}_\mathrm{env}$--$M_\mathrm{disk}$ space in Fig.~\ref{fig:S08_parameters}, rules out the option that S8 is an embedded Stage I source. The relative disk mass is high enough to rule out Stage III and S8 is therefore classified as a Stage II object.  

Objects S11, S14 and S16 are all intrinsically faint in the IRAC bands. It is evident from the SED plots of these objects in Fig.~\ref{fig:YSOSEDs} that model fitting therefore benefits less from upper limits from MIPS~70~$\mu$m and SCUBA 450~$\mu$m and 850~$\mu$m. The lack of data points and stringent upper limits at far-infrared and sub-millimeter wavelengths results in an ambiguity: the presence of the bump longward of $\sim$30~$\mu$m, a characteristic feature in SEDs of Stage I sources, cannot be ruled out. This is reflected by the envelope accretion rate in column (6) of Table \ref{table:allYSOs}, which ranges from 0 to well above the $\dot{M}_\mathrm{env}=10^{-6} M_\star\mathrm{yr}^{-1}$ threshold. For S14, the majority of models that obey the criterion in Eq.~\ref{eq:chi2criterion} use an envelope accretion rate below this threshold, and it is therefore classified to be most likely a Stage II object, with a mass between 0.5 and 4~$M_\odot$. S11 and S16, however, show a particularly large spread of models covering both Stage I and Stage II parameters, which is why their evolutionary stage in Table~\ref{table:allYSOs} is listed as ``Stage I or II". The models fitted to S11 and S16 have stellar masses ranging from 0.1 to 5 and from 2 to 5~$M_\odot$, respectively. 

\subsubsection{S13 and S20}
\label{sec:uncertainobjects}
The criterion in Eq.~\ref{eq:chi2criterion} leaves only one YSO model for S13, and even this is a relatively poor fit (see Fig.~\ref{fig:YSOSEDs} and column (4) of Table~\ref{table:allYSOs}). Hence, no firm conclusion can be drawn regarding any of the key physical parameters such as stellar mass, disk mass, envelope accretion rate, and therefore its evolutionary stage. 

Finally, S20 is the only source classified as YSO which is not spatially (in the plane of the sky) associated with the IRDC. Its key parameters are poorly constrained due to the high deviation between the best-fitting model and the data points of S20 (see Fig.~\ref{fig:YSOSEDs}). Therefore, the derived parameters and evolutionary classification of this objects should be considered with care.

\subsection{Color Analysis of S1 and S3}
\label{sec:colorcolor}
Since S1, S3 and S10 have less than 3 data points, it is noted in Sect.~\ref{sec:modelfitting} that model SEDs cannot be fit to these sources. Source S10 is overwhelmed by diffuse background radiation. We are therefore reluctant to draw conclusions about the physical properties of this object based on its two data points. 

Sources S1 and S3 on the other hand, although they only have a flux measurement at 24~$\mu$m, can be examined purely by their $[8.0]-[24]$ color. Using the MIPS zero point magnitudes from the SSC website\footnote{{\tt http://ssc.spitzer.caltech.edu/mips/calib/}}, S1 and S3 have a  [24] magnitude of 6.1 and 5.8, respectively. Using the 8.0~$\mu$m upper limits from Table~\ref{table:sourcelist} and the IRAC magnitude system defined in \citet{reach2005}, the [8.0] magnitude is at least 11.3 for each source. This puts a lower limit to their $[8.0]-[24]$ colors: $>5.2$ for S1 and $>5.5$ for S3. Figure~23 from \citet{robitaille2006} shows that practically only Stage I YSO models exhibit an $[8.0]-[24]$ color this red. In conclusion, if an evolutionary phase is to be attributed to sources S1 and S3 based on the above argument, it would be Stage I. 

\section{Conclusions and Discussion}
\label{sec:conclusions_and_discussion}

\subsection{Conclusions}
\label{sec:conclusions}

Of the 20 sources near IRDC G48.65 that are visible at 24\,$\mu$m, a total of 13 are classified as YSOs, seven of which are found to be in Stage I, two are in Stage II, two more are in either Stage I or Stage II, and two are uncertain. While all reliably classified YSOs lie along the IRDC extinction filament, the four sources matched to photosphere models all lie away from the filament. Each YSO generally resides within a projected distance of $<$1\,$\mathrm{pc}$ from another YSO. We conclude that G48.65 is an example of a dark cloud environment forming a group of stars.

The most important modeled properties of the 24\,$\mu$m sources that are classified as YSOs are summarized in Table \ref{table:allYSOs}. The stellar masses of the modeled objects range from slightly sub-solar to $\sim$8\,$M_\odot$. The inferred evolutionary Stages are predominantly early phases, ranging from early Stage I to late Stage II. This is believed to be caused partly by the selection effect discussed in Sect.~\ref{sec:MIPSselection}. Therefore, the presented range of evolutionary phases should not be interpreted as a full picture of all star forming activity in this IRDC. 

The emission peak P1 identified by \citet{ormel2005} is resolved by Spitzer into two distinct emission cores: S6 and S5. Both objects are found to be in the earliest Stage of star formation. Stellar masses are not well constrained for these sources but are unlikely to be in excess of $8\,M_\odot$. The summed total luminosity of the best-fit models for S5 and S6 is $\sim$$10^2$--$10^3\,L_\odot$, consistent with results from modeling by \citet{ormel2005} based purely on sub-mm observations. The total luminosity of the best fitting model for objects S12 (EP) and S15 (P2) are consistent with values found by \citet{ormel2005}.

\subsection{Robustness of Results}
\label{sec:discussion}
It is important to assess a YSO based on data points in a wavelength regime as extended as possible. While a detection of a certain YSO at sub-millimeter wavelengths seems to be a strong indication that it is in Stage I (S5, S6, S12, S15), a detection by 2MASS in the near-infrared does not rule out a Stage I classification, as sources S2 and S7 show. It may therefore lead to erroneous classification of YSOs if one bases it either on only near-infrared or on only far-infrared/sub-millimeter data. 

The availability of data points in the 3--30\,$\mu$m regime is in some cases enough to constrain the evolutionary stage of a YSO. Source S8 (see Fig.~\ref{fig:YSOSEDs} for the SED fit) is an example in which the degeneracy of the model fitting is decreased considerably by adding 2MASS data points (see also Sect.~3.3 and Fig.~3 of \citet{robitaille2007}).

\subsection{Continuous Low-mass Star Formation}

The Stage I sources in particular show large uncertainties in stellar mass (see Table~\ref{table:allYSOs}). This is not surprising considering that, for objects that emit mostly through re-radiation from the accreting envelope, it is of course inherently difficult to model parameters of the obscured central heating source. However, the OH maser survey by \citet{pandian2007} indicates the absence of OH maser emission in the direction of IRDC G48.65. Since OH masers are commonly associated with high-mass star formation ($\gtrsim$8\,$M_\odot$), this agrees with the fact that we find no massive YSOs in G48.65. Considering the above, we are confident that any YSO associated to G48.65 is unlikely to harbor a central star more massive than 8\,$M_\odot$. 

Based on the range of stellar masses and evolutionary Stages of the YSOs, we conclude that G48.65 has been an active low-mass star forming region for at least $\sim$$10^6$ years (late Stage II) up to as recent as $\sim$$10^4$ years ago (early Stage I) and is likely to be in the process of forming still younger stars. Assuming that the IRDC is indeed the birth place of these young stars, the cloud itself must have been stable over a time scale of at least $\sim$$10^6$ years. This is consistent with modern views on lifetimes of $\lesssim$$10^6$ years for molecular clouds \citep[as reviewed in][]{maclow2004} and with lifetimes of $\lesssim$$10^7$ years for dense cluster forming clouds \citep[see review by][]{larson2003}.

\subsection{Mid-infrared Selection Effect} 
\label{sec:MIPSselection}
The starting point of the study in this paper -- the set of point sources in the MIPS 24\,$\mu$m image -- introduces an observational bias toward the younger evolutionary stages. Early stage YSOs are simply brighter in the mid-infrared, since the total mid-infrared luminosity of a YSO generally decreases with time. Especially in a fairly distant region such as IRDC G48.65, at $\sim$2.5\,kpc, one must keep in mind that later stages and intrinsically fainter sources are less likely to be detected. This study in particular is biased toward the earlier stages, since it focusses on the 24\,$\mu$m objects. 

It is primarily the relatively cold envelope component, dominant in Stage I objects, that gives a YSO SED its large ``bump" longward of 20~$\mu$m. Hence, an analysis of objects near the IRDC that are only visible at wavelengths shortward of 8~$\mu$m is expected to lead to the identification of more evolved (Stages II and III) and less massive YSOs. The IRAC 8~$\mu$m image, for example, shows at least a factor 3 more point sources than the 24~$\mu$m image in the same field of view. Such a study will be essential in getting a complete census of the star formation activity in G48.65, both in terms of time scales and in terms of a stellar mass function. 

In principle, the set of young stars of various masses presented in this paper could be used to construct an initial mass function (IMF). Extrapolation of such an IMF could provide insight into the absence of massive stars ($>$8\,$M_\odot$). However, we refrain from presenting such an exercise here. The first reason for this is the incompleteness of our sample discussed above; a second reason is that the inferred stellar mass of each individual YSO (see Col.~(5) of Table~\ref{table:allYSOs}) is rather poorly constrained.

\subsection{Outlook}
\label{sec:outlook}

In the near future, the 2MASS point source catalog (1--3\,$\mu$m) and the GLIMPSE catalog (3--9.5\,$\mu$m) will be complemented by the MIPSGAL point source list of the Galactic plane at 24 and 70\,$\mu$m. Moreover, the UKIRT Infrared Deep Sky Survey \citep{lawrence2007} is already partly available to the general community and will eventually cover half of the Galactic plane in the $J$, $H$ and $K$ bands, superseding the sensitivity of 2MASS by roughly 3 magnitudes. This will open up the possibility to investigate YSOs in hundreds of other IRDCs in the Galaxy using an approach similar to the one taken in this paper. This approach has shown that source-by-source investigation of emission cores with 1--30\,$\mu$m data can lead to the identification of young stars in various phases of evolution. Perhaps positions that show sub-millimeter emission are the most interesting cases to start with if one is attempting to identify very early stages of star formation. However, even with the combination of data from 2MASS, GLIMPSE and MIPSGAL, covering just the 1--70\,$\mu$m regime, it is possible to systematically study star formation activity across the Galactic plane. IRDCs may eventually be divided into groups according to whether they show star formation activity, and if they do, whether they form low-mass stars or high-mass stars.

The Stage I sources in Table \ref{table:allYSOs} are expected to show associated outflows and shocks, traced by e.g. SiO. Detection of such tracers could confirm the star forming activity and reveal kinematic and spatial structure of YSOs. In addition, measurements by the VISIR instrument (8--13\,$\mu$m) at VLT will be able to resolve more detailed structure ($<$$10^3\ \mathrm{AU}$ at 2.5\,kpc) of the sources and show envelope and disk components. It may even show that objects that appear as one source in the Spitzer images are in fact multiple YSOs. Source S5, for example, appears slightly elongated in the Spitzer images. In fact, some of the poorer model fits (see Col.~(4) in Table~\ref{table:allYSOs}) may be explained by the fact that the models do not account for multiple young star-disk-envelope systems inside one telescope beam. 

The high spatial and spectral resolution and new wavelength regimes of future observatories, such as ALMA, the Herschel Space Observatory \citep[specifically the Hi-GAL survey,][]{noriega-crespo2008} and the James Webb Space Telescope, will enable more detailed studies of individual objects associated with G48.65 and other (distant) IRDCs.

\begin{acknowledgements}
The authors are grateful to Thomas Robitaille for his help with the use of the YSO models, and for making adjustments to the fitting tool upon request. Furthermore, the authors wish to thank Floris van der Tak and Marco Spaans for their support and suggestions that were essential for the interpretation of the results and the completion of the manuscript. We thank Chris Ormel and Wilfred Frieswijk for additional constructive discussions, comments and suggestions that helped to improve this paper. We acknowledge the efforts of Kathleen Kraemer for preparing the observing proposal for the \textit{Spitzer} data on which this paper is based and for her comments on the manuscript. We thank the anonymous referee whose comments and suggestions helped to improve the content as well as the clarity and presentation of this paper.

This work is based on observations obtained with the \textit{Spitzer Space Telescope}, which is operated by the Jet Propulsion Laboratory, California Institute of Technology, under NASA contract 1407. In addition, this publication makes use of data products from the Two Micron All Sky Survey, which is a joint project of the University of Massachusetts and the Infrared Processing and Analysis Center/California Institute of Technology, funded by the National Aeronautics and Space Administration and the National Science Foundation.
\end{acknowledgements}

\bibliographystyle{bibtex/aa}  
\bibliography{references}

\end{document}